\documentclass[pre,twocolumn]{revtex4}

\usepackage{amsmath}
\usepackage{amssymb}

\usepackage{graphicx}

\newcommand{\kbar}{\mathchar'26\mkern-9mu k}

\begin{document}

\title{Optimal Coordinate as a General Method in Stochastic Dynamics.}

\author{Sergei V. Krivov}

\affiliation{Astbury Center for Structural Molecular Biology, University of Leeds, Leeds, United Kingdom}

\begin{abstract}
A general method to describe stochastic dynamics of Markov processes is suggested. The method aims to solve three related problems. The determination of an optimal coordinate for the description of stochastic dynamics.  The reconstruction of time from an ensemble of stochastic trajectories. The decomposition of stationary stochastic dynamics on eigen-modes which do not decay exponentially with time. The problems are solved by introducing additive eigenvectors which are transformed by a stochastic matrix in a simple way - every component is translated on a constant distance. Such solutions have peculiar properties. For example, an optimal coordinate for stochastic dynamics with detailed balance is a multi-valued function. An optimal coordinate for a random walk on the line corresponds to the conventional eigenvector of the one dimensional Dirac equation. The equation for the optimal coordinate in a slow varying potential reduces to the Hamilton-Jacobi equation for the action function.
\end{abstract}

\maketitle

\section*{Introduction}
The description of a complex, multidimensional, stochastic process is often simplified by projecting it on one or a few variables \cite{du_transition_1998,best_reaction_2005, coifman_diffusion_2006,  nadler_diffusion_2006, altis_dihedral_2007, krivov_diffusive_2008, krivov_optimal_2011, krivov_free_2011, peters_reaction_2013}. During such dimensionality reduction one may lose some information; hence the variables should be optimally selected to preserve the information of interest. Here we are interested in the selection of variables that preserve the information about the dynamics.
One aims to select a coordinate such that the dynamics projected on the coordinate is 
approximately Markovian, i.e., it is independent from the dynamics along other degrees of freedom. In other words, the current value of the coordinate alone determines the future dynamics of the coordinate. Such dynamics are often described as diffusion on a free energy profile with a position dependent diffusion coefficient, which can be determined from the coordinate time series \cite{krivov_reaction_2013}. 

The folding (splitting or committor) probability is considered to be an optimal coordinate \cite{vanden-eijnden_assumptions_2008, krivov_reaction_2013, berezhkovskii_diffusion_2013} for the description of transition dynamics between any two chosen boundary states, i.e., a reaction. The name comes from the protein folding field, where this coordinate is equal to the probability of reaching the folded state before reaching the unfolded state starting from the current configuration \cite{du_transition_1998}. The projection on the coordinate preserves some properties of the dynamics, in particular, the equilibrium flux between the boundary states, and the committor probability, by construction. These properties can be computed exactly by simulating diffusive dynamics with the determined free energy landscape and diffusion coefficient \cite{krivov_reaction_2013}. 

The coordinate, however is exact only for the description of the equilibrium transition dynamics between two boundary states. The dynamics inside the boundary states, or dynamics in general, without defining two end states, can not be described.
It may seem unlikely that a single coordinate, even though optimally selected, can give a complete description of multidimensional complex dynamics. Classical mechanics, however, provides an example. The action or the Hamilton's principal function  $S(x_i,t)$ gives complete description of the dynamics of a system governed by the deterministic equations of classical mechanics. Here we suggest a class of optimal coordinates for the description of  stochastic dynamics in general.  We show that under some conditions the equations for the optimal coordinate, suggested here, are reduced to the Hamilton-Jacobi equation for the action function. In other words, the suggested optimal coordinate can be considered as a generalization of the action function to stochastic dynamics. The problem of the determination of such an optimal coordinate is closely related to two other problems.

\textbf{The eigen-modes for stochastic dynamics.}
The decomposition of the dynamics of a multidimensional harmonic oscillator on normal modes is an example of a Markovian projection on coordinates. Each mode evolves independently. The phase of each normal mode can be considered as an optimal coordinate. A quantum mechanical wave function, which is a linear combination of basis eigen-functions, is another example. Can one introduce analogous concepts for equilibrium stochastic dynamics? The conventional decomposition of the probability distribution on the eigenvectors of the master equation is not appropriate. Since all the eigenvalues (but the first) have negative real part \cite{risken_fokker-planck_1996} the projection on each eigenvector exponentially decay with time. Thus after a finite amount of time only the equilibrium eigenvector survives. The latter does not describe dynamics. However, if one observes a particular dynamical trajectory of the process, the dynamics becomes stationary but never stops. Which leads us to the second problem. Can one define such eigen-modes that can be used to describe stationary stochastic dynamics. 

The folding probability optimal coordinate which monitors progress of the folding reaction, increases as the system comes closer to the folded state. It is natural to expect that an optimal coordinate that monitors progress of the dynamics in general, without any relation to boundary states, steadily increases. In particular, it should increase whenever the system changes its state. A variable which always increases, whenever the system changes its state is time. Which leads us to the third problem.

\textbf{The reconstruction of time.}
Assume that we observe a stochastic process, generated by an unknown transition probability or transition rate matrix. We have access to all the variables representing the state of the system apart from the time variable (which is external to the system). For example, one is given a trajectory of the system sampled with random unknown time intervals. Can one reconstruct the time variable? Such reconstruction can be useful, for example, if one wants to determine the transition probability matrix.

Let $W(x)$ be such a function of coordinates that can be used to reconstruct the time interval as $t_2-t_1\sim W(x(t_2))-W(x(t_1))$, where $x(t)$ is a trajectory. Since the dynamics is stochastic, such estimates fluctuate around a true value. Thus, to determine time accurately, one needs to average it over an ensemble of trajectories. The time interval can be estimated more precisely 
as $t_2-t_1\sim 1/N\sum_\alpha [W(x_\alpha(t_2))-W(x_\alpha (t_1))]$, where the average is taken over an ensemble of trajectories $x_\alpha (t)$ ($\alpha=1,N$) leading from an initial distribution $x_\alpha(t_1)$ to a final  distribution $x_\alpha(t_2)$ and $N\rightarrow \infty$.   

Any such function that allows accurate time reconstruction can be considered as an optimal coordinate. The trajectory projected on such a coordinate has simple dynamics. There is no need to compute the free energy profile and the diffusion coefficient. Starting from the current position $t$,  its position after a time interval $\Delta t$ is equal (on average) to $t+\Delta t$, i.e.,  it depends only on the current position. 

While the optimal coordinate $W(x)$ describes the stochastic dynamics in a simple way, it might be useful to be able to map this description back to the original dynamics. In principle, one can invert the relationship. Given $x_\alpha(t_1)$ and $t_2-t_1$, one may attempt to determine $x_\alpha(t_2)$. Since we have just a single equation to determine the final distribution $x_\alpha(t_2)$, the problem is ill-defined. It can, however, be solved in the following cases. The first case, when one is interested in a single parameter of the distribution, for example, an average of some operator like the mean position. The time dependence of a single parameter can be determined from the single equation. The second case, if the initial distribution is an eigen mode of the dynamics, then (by construction) the distribution does not change with time. The only changing parameter is the "phase" which can be determined from the single equation. The general solution is then obtained as a superposition of all such eigen-modes. This case corresponds to the conventional way of solving a linear equation by decomposing it onto a sum of eigen-modes, i.e., it provides the solution to the second problem.

Here we introduce a general method to solve the three problems. Briefly, the main difference between the proposed method and the conventional one is to seek the solution of the master equation in the form $S=W-\nu t$, instead of conventional $S=\psi e^{\lambda t}$. The new solution has a number of interesting, peculiar and counter-intuitive properties.  For example, the optimal coordinate is a multi-valued function. To familiarize the reader with the new concepts we extensively use illustrative examples. We start by deriving the equations for the optimal coordinate by requiring it to be an ideal clock.

\section*{Optimal coordinate as an ideal clock}
\textbf{Equilibrium optimal coordinate}. To illustrate counter intuitive properties of the optimal coordinate we first consider a more straightforward case of an equilibrium optimal coordinate. Consider an \textbf{ideal system} where a point performs a random walk along x with constant a diffusion coefficient and zero mean displacement. In this case the mean square displacement grows with time as $\langle \Delta x^2(\Delta t)\rangle=2D\Delta t$. If one is given snapshots of the position of the point (trajectory) $x(t_i)$, one may estimate the time intervals between the snapshots (reconstruct the time) as $t_{i+1}-t_i=\Delta t\sim \Delta x^2/2D=[x(t_{i+1})-x(t_i)]^2/2D$. Since the process is random such estimate fluctuates around the true value. To improve the accuracy one may consider an ensemble of identical systems. Given an ensemble of trajectories $x_\alpha(t)$ ($\alpha=1,N$) sampled at the same (unknown) time points $t_i$, the time interval between the snapshots can be reconstructed with arbitrary accuracy as 
\begin{multline*}
t_{i+1}-t_i=\Delta t=\langle \Delta x^2\rangle_\alpha/2D=\\1/(2DN) \sum_\alpha [x_{\alpha}(t_{i+1})-x_{\alpha}(t_i)]^2
\end{multline*}
For a \textbf{real system} where the diffusion coefficient or the potential energy surface depends on the coordinate, $\langle \Delta x^2\rangle$ does not grow strictly linear with time. However, for any such system one can find a coordinate $W$ (an optimal coordinate), so that the mean square displacement of the coordinate $\langle \Delta W^2\rangle$, computed for an equilibrium ensemble of trajectories, grows linearly with time \cite{krivov_reaction_2013}.

Conversely, define the optimal coordinate as such coordinate whose mean square displacement grows linearly with time. Let us introduce some notation \cite{krivov_reaction_2013}. Consider a Markov process with transition matrix $\textbf{P}$, where $P_{ji}(\Delta t)$ is the probability of transition from state $i$ to $j$ after time interval $\Delta t$
\begin{equation}
P_i(t+\Delta t)=\sum_j P_{ij}(\Delta t) P_j(t).
\label{markov}
\end{equation}
The transition probability matrix for time interval $n \Delta t$ is  $\textbf{P}(n\Delta t)=\textbf{P}^n(\Delta t)$. 
Consider a stationary (steady-state) ensemble of trajectories $x_\alpha(t)$ ($\alpha=1,N$), generated by a Markov process (Eq. \ref{markov}). We assume that the configuration space of the system is discrete and is represented by a (possibly infinite) set of integer numbers, i.e., indexes. If the original system's dynamics are defined in a continuous configuration space, we assume that the space has been discretized. Thus, each trajectory $x_\alpha(t)$ is just a sequence of such indexes denoting current state. Such a representation is manifestly invariant with respect to the choice of the coordinate system. If trajectories are sampled with a constant time interval $\Delta t$ one can determine the transition matrix $n_{ji}(\Delta t)$, which equals the number of transitions from state i to state j. $n_i=\sum_j n_{ji}=\sum_j n_{ij}$ is the number of times state i has been visited, which is proportional to $P^{st}_i$, the stationary (steady-state) probability distribution $P^{st}_i=\sum_j P_{ij}(\Delta t)P^{st}_j.$ Based on $n_i$ and $n_{ji}(\Delta t)$, the transition probability matrix can be estimated as $P_{ji}(\Delta t)=n_{ji}(\Delta t)/n_i$. Let superscript T denote properties associated with the ensemble of time-reversed trajectories, i.e., trajectories are read in opposite direction, from the end to the start.
These trajectories can be considered as a realization of a Markov process with  
$P_{ji}^T(\Delta t)=n^T_{ji}(\Delta t)/n_i$, where $n^T_{ji}(\Delta t)=n_{ij}(\Delta t)$ and $n^{T}_i=n_i$ \cite{norris_markov_1998}.


If W is such that for every $i$ 
\begin{equation}
\sum_jP_{ji}(\Delta t) (W_j-W_i)=0,
\label{rceq}
\end{equation}
then
\begin{equation}
\langle\Delta W^2(n\Delta t)\rangle=n\langle\Delta W^2(\Delta t)\rangle=2Dn \Delta t.
\label{dw2}
\end{equation}
 We prove by induction. Assume that the statement is valid for n, then
\begin{align*}
\langle\Delta W^2((n+1)\Delta t)\rangle=&\sum_{ij} P_{ji}(n\Delta t+\Delta t)P^{st}_i(W_j-W_i)^2=\\\sum_{ijk} P_{jk}(\Delta t)P_{ki}(n\Delta t)&P^{st}_i(W_j-W_k+W_k-W_i)^2=\\\sum_{ijk} P_{jk}(\Delta t)P_{ki}(n\Delta t)&P^{st}_i[(W_j-W_k)^2+\\2(W_j&-W_k)(W_k-W_i)+(W_k-W_i)^2]=\\
\sum_{jk} P_{jk}(\Delta t)P^{st}_k(W_j&-W_k)^2+\\
2\sum_{ik} P_{ki}(n\Delta t)P^{st}_i(&W_k-W_i)\sum_jP_{jk}(\Delta t)
(W_j-W_k)+\\\sum_{ik} P_{ki}(n\Delta t)P^{st}_i&(W_k-W_i)^2=\\ &\langle\Delta W^2(\Delta t)\rangle+n\langle\Delta W^2(\Delta t)\rangle
\end{align*}
Analogously, from Eq. \ref{rceq} it follows that for all $n$
\begin{equation}
\sum_jP_{ji}(n\Delta t) (W_j-W_i)=0,
\label{npfold}
\end{equation}
i.e., the optimal coordinate is the same for the dynamics, sampled with a different constant sampling interval. We prove by induction. Assume that $\sum_jP_{ji}(n\Delta t) (W_j-W_i)=0$, then
\begin{multline*}
\sum_{j} P_{ji}(n\Delta t+\Delta t)(W_j-W_i)=\\\sum_{jk} P_{jk}(\Delta t)P_{ki}(n\Delta t)(W_j-W_k+W_k-W_i)=\\
\sum_k P_{ki}(n \Delta t) \sum_{j} P_{jk}(\Delta t)(W_j-W_k)+\\
\sum_{k} P_{ki}(n\Delta t)(W_k-W_i)=0
\end{multline*}

The transition matrix for a trajectory sampled with random intervals is the average
$\langle P_{ij}\rangle=\sum_n \rho(n) P_{ij}(n \Delta t)$, where $\rho(n)$ is the probability of having interval of $n \Delta t$. Averaging Eq. \ref{npfold} with $\rho(n)$ one finds that the optimal coordinate can be found from
\begin{equation}
\sum_j \langle P_{ji} \rangle (W_j-W_i)=0.
\label{rceqvardt}
\end{equation}
In summary, given a stationary ensemble of trajectories $x_\alpha(t)$ ($\alpha=1,N$), sampled at unknown time points $t_i$, one can determine the averaged transition matrix $\langle P_{ji} \rangle$ and thus the optimal coordinate $W$ with Eq. \ref{rceqvardt}. Using the optimal coordinate the time interval between two time points can be reconstructed (up to a constant factor determined by D)
\begin{multline}
t_j-t_i=\langle \Delta W^2 \rangle_\alpha/2D=\\1/(N2D)\sum_{\alpha=1,N} [W_{x_\alpha(t_j)}-W_{x_\alpha(t_i)}]^2.
\label{seqdt}
\end{multline}
Here $W_{x_\alpha(t_j)}$ denotes the value of the optimal coordinate $W_i$ at state $i=x_\alpha(t_j)$, which is attained by trajectory $\alpha$ at time instant $t_j$.
Note that given both direct and time-reversed trajectories, Eq. \ref{seqdt} predicts only increase in time, which is in agreement with equilibrium statistical mechanics, where there is no difference between forward and time-reversed processes.

\textbf{The optimal coordinate can have neither a maximum nor a minimum.} The equation for the optimal coordinate (eq. \ref{rceq}) can be satisfied for every $i$, only if for every $i$  there are such $j$ that $W_j<W_i$ and such $j$ that $W_i<W_j$ because $P_{ji}> 0$. For systems with infinite configuration space this does not seem to be a problem, e.g., for a random walk on the (infinite) line $W=x$, whereas systems with finite configuration space require special consideration, because they have a finite set of values of $W_i$ and hence have maximum and minimum $W_i$. Consider a random walk on a ring, with probability 1/2 to jump left or right. The transition matrix is $p_{i,i+1}=p_{i,i-1}=p_{1,N}=p_{N,1}=1/2$. Consider the optimal coordinate as a function of angle $\phi$ for $\phi=[0,2\pi)$. Then the equation for the optimal coordinate is $W_{i}-W_{i-1}=W_{i+1}-W_i$, which means that points $W_i$ are placed equidistantly on the ring $W_{i+1}-W_i=\mathrm{const}=2\pi/N$. If one starts from $W_1=0$ and uses the equation to consequently determine $W_{i+1}$ from $W_i$ along the ring, then when one completes the loop and returns to the first node one obtains $W_1=2\pi$. After the second loop $W_1=4\pi$ and so on. Thus, to satisfy Eq. \ref{rceq} for all $i$, \textbf{the optimal coordinate has to be a multi-valued function}. For the ring  $W=\phi$, for $\phi=(-\infty, \infty)$ is the phase angle that covers the ring periodically. The inverse function, the mapping from the optimal coordinate to the states, is periodic. 

Eq. \ref{rceq} can be rewritten for a single-valued function $W_i$ restricted to any branch, as
\begin{equation}
\sum_jP_{ji}(\Delta t) (W_j+d_{ji}-W_i)=0,
\label{rceqd}
\end{equation}
where $d_{ji}$ denotes the increment in the coordinate between two branches of the multi-valued function. For a random walk on a ring $d_{1,N}=-d_{N,1}=2\pi$ and otherwise $d_{ji}=0$. Eq. \ref{rceqd} is the conventional system of linear equations on a single valued function, and can be solved by linear algebra methods. Note that, since the equation defines the solution up to a constant $W_i=W_i+c$, to solve it on a computer, one should supplement it with an equation which fixes the constant, for example, $W_1=0$.

\begin{figure}[htbp]
\centering 
\resizebox*{\columnwidth}{!}{\includegraphics*[]{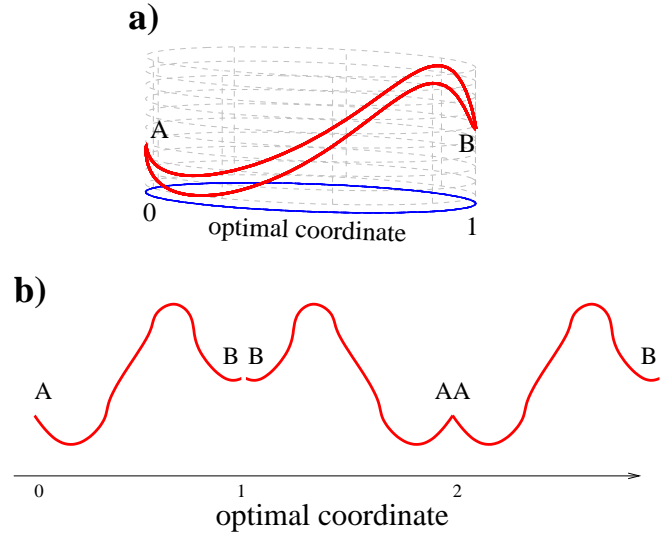}}
\caption{a) Construction of an optimal coordinate with ring topology by joining $p_{fold}$ coordinate and its mirror image at the boundaries. b) The resulting (periodic) profile along an infinite periodic coordinate of the ring.  For such coordinate Eq. \ref{dw2} is valid for all n. The red lines show a model free energy profile.}
\label{pfold0110}
\end{figure}

Similar construction can be made for the dynamics on a segment between two boundary states A and B. Using the folding probability ($p_{fold}$) as an optimal coordinate the segment is mapped onto the $[0,1]$ segment, so that Eq. \ref{rceq} is satisfied for all the points but A and B, which are mapped to $0$ and $1$, respectively \cite{krivov_reaction_2013}. To make the equation valid at points A and B, the $[0,1]$ segment and its mirror copy $[1,0]$ are joined together to form a ring,  $0$ ends are joined together and $1$ ends are joined together. Fig. \ref{pfold0110}a visualizes the construction as a drawing on the surface of a cylinder; the joint profile wraps the cylinder. Eq. \ref{rceq} is satisfied for nodes A and B due to symmetry. Fig. \ref{pfold0110}b shows a schematic realization of the mirroring construction along an infinite periodic optimal coordinate of the ring. For such a coordinate Eq. \ref{dw2} is valid for all n. A practical realization of  the procedure during an analysis of a reaction coordinate time-series is as follows. Whenever the system reaches either A or B, a new current branch is selected out of the two with equal probability of 0.5. An alternative way to make Eq. \ref{dw2} valid is to modify the counting scheme by considering the transition paths \cite{krivov_reaction_2013}, which is not discussed here.

On this we finish the discussion of the equilibrium optimal coordinate and switch to  a more powerful method which can be applied to non-equilibrium ensembles of trajectories and can estimate the change of time in both positive and negative directions.

\textbf{Non-equilibrium optimal coordinate.} 
Consider \textbf{an ideal system} where a point performs random jumps to the right with distance $a$ and rate $r$. In this case the average distance the system transits during time $\Delta t$ is $\Delta x=ra \Delta t$. Accordingly, the time interval between two snapshots of the trajectory separated by distance $\Delta x$ can be estimated as $\Delta t=\Delta x/(ra)$. For a realistic system, where the rate and jump distance can vary, $\Delta t=\Delta x/(ra)$ is no longer valid. Again, for any system an optimal coordinate $W$ can be constructed so that time intervals can be determined as $\Delta t=\langle\Delta W\rangle/\nu$, where $\nu$ is a constant, with dimension of frequency; W is dimension-less. 

\textbf{Left additive eigenvector.}
Let $W^L$ and $\nu$ be a solution of 
\begin{equation}
\sum_{j}  n_{ji}(\Delta t)(W^L_j-W^L_i -\nu \Delta t)=0,
\label{wldef2}
\end{equation}
or
\begin{equation}
\sum_{j} P_{ji}(\Delta t)(W^L_j-W^L_i -\nu \Delta t)=0,
\label{wldef}
\end{equation}
which can be considered as the definition of the left \textit{additive} eigenvector
\begin{equation}
\sum_{j} P_{ji}(\Delta t)W^L_j=W^L_i +\nu \Delta t,
\label{levec}
\end{equation}
where $\nu \Delta t=\lambda$ is an \textit{additive} eigenvalue. For a system with $n$ states Eq. \ref{wldef2} consists of n equations, which together with the equation that fixes the origin of the eigenvector (e.g., $W^L_1=0$) makes it $n+1$ equations for $n+1$ variables. The multiplication by (the transpose of) matrix \textbf{P} changes the components of the vector $\mathbf{W}^L$ in a simple way by adding a constant.
It is easy to see that
\begin{equation*}
\sum_{j} P_{ji}(n\Delta t)W^L_j=\sum_j P^n_{ji}(\Delta t)W^L_j=W^L_i +n\nu \Delta t
\end{equation*}
or
\begin{equation}
\sum_{j} P_{ji}(n\Delta t)(W^L_j-W^L_i -n\nu \Delta t)=0.
\label{wlndt}
\end{equation}
For example, for n=2:
\begin{align*}
\sum_{jk} P_{jk}(\Delta t)P_{ki}(\Delta t)W^L_j=\sum_{k} P_{ki}(\Delta t)(W^L_k+\nu \Delta t)=\\
\sum_{k} P_{ki}(\Delta t)W^L_k+\sum_{k} P_{ki}(\Delta t)\nu \Delta t=
W^L_i+2\nu \Delta t.
\end{align*}
If the transition matrix $\langle P_{ji}\rangle=\sum_n \rho(n) P_{ji}(n \Delta t)$ is the average of the transition matrix with random distribution of steps $\rho(n)$ (a trajectory sampled with random intervals), then $W^L$ is also the solution of
\begin{equation}
\sum_{j} \langle P_{ji}\rangle (W^L_j-W^L_i -\langle \Delta t\rangle \nu)=0,
\label{wlrandt}
\end{equation}
where $\langle \Delta t\rangle=\sum_n \rho(n) n \Delta t$ is the average sampling interval.
Multiplying Eqs. \ref{wlndt} and \ref{wlrandt} by $n_i$ one obtains
\begin{equation}
\sum_{j} n_{ji}(n\Delta t)(W^L_j-W^L_i -n\nu \Delta t)=0,
\label{wlndt2}
\end{equation}
and
\begin{equation}
\sum_{j} \langle n_{ji}\rangle (W^L_j-W^L_i -\langle \Delta t\rangle \nu)=0.
\label{wlrandt2}
\end{equation}
Thus, given an ensemble of trajectories $x_\alpha(t)$ ($\alpha=1,N$), sampled at unknown time points $t_i$, one can determine the averaged transition matrix $\langle n_{ji} \rangle$ and thus the optimal coordinate $W^L$ with Eq. \ref{wlrandt2}.

\textbf{Right additive eigenvector.} It is useful to define the right \textit{additive} eigenvector as a solution of equation
\begin{equation}
\sum_{j} n_{ij}(\Delta t)(W^R_i-W^R_j -\nu \Delta t)=0
\label{wrdef}
\end{equation}
or, equivalently
\begin{equation}
\sum_{j}  \tilde{P}_{ij}(\Delta t)(W^R_i-W^R_j -\nu \Delta t)=0
\label{wrdef2}
\end{equation}
and
\begin{equation*}
\sum_{j}  \tilde{P}_{ij}(\Delta t)W^R_j=W^R_i -\nu \Delta t,
\end{equation*}
where $\tilde{P}_{ij}(\Delta t)=n_{ij}(\Delta t)/n_i=P_{ij}(\Delta t)P^{st}_j/P^{st}_i$.
It is easy to see that 
\begin{equation*}
\sum_{j}  \tilde{P}_{ij}(n\Delta t)(W^R_i-W^R_j -n\nu \Delta t)=0.
\end{equation*}
and 
\begin{equation}
\sum_{j}  n_{ij}(n\Delta t)(W^R_i-W^R_j -n\nu \Delta t)=0
\label{wrndt2}
\end{equation}
Note that $\tilde {P}_{ij}$ is not a stochastic matrix, i.e., $\sum_i \tilde{P}_{ij}\neq 1$, however $\sum_j \tilde{P}_{ij}= 1$. If detailed balance holds, i.e., $n_{ji}(\Delta t)=P_{ji}(\Delta t)n_i=P_{ij}(\Delta t)n_j=n_{ij}(\Delta t)$ then $\tilde{P}_{ij}=P_{ji}$. 

Given an ensemble of trajectories $x_\alpha(t)$ ($\alpha=1,N$), which describes stationary dynamics of the system, one can define the following averages to measure time intervals. Averaging over the entire ensemble of trajectories
\begin{equation}
1/N\sum_\alpha [W_{x_\alpha(t_2)}-W_{x_\alpha(t_1)}] 
\label{ensall}
\end{equation}
Averaging over the subset of trajectories starting from a particular state at time $t_1$ (or a subset of states)
\begin{equation}
\frac{\sum_\alpha [W_{x_\alpha(t_2)}-W_{x_\alpha(t_1)}]A_{x_\alpha(t_1)}}{\sum_\alpha A_{x_\alpha(t_1)}}
\label{ensstart}
\end{equation}
where A is the indicator function of the subset of states, i.e., $A_x=1$ if x is in the chosen subset of states and zero otherwise. For a single state $i$, $A_x=\delta_{xi}$, the Kronecker symbol.
Averaging over the subset of trajectories ending in a particular state at time $t_2$ (or a subset of states)
\begin{equation}
\frac{\sum_\alpha [W_{x_\alpha(t_2)}-W_{x_\alpha(t_1)}]A_{x_\alpha(t_2)}}{\sum_\alpha A_{x_\alpha(t_2)}}.
\label{ensend}
\end{equation}
Eqs. \ref{ensstart} and \ref{ensend} reduce to Eq. \ref{ensall}  for $A_x=1$ for all x.

Multiplying Eq. \ref{wlndt2} by $A_i$ and summing over $i$ one finds that \textbf{the left eigenvector can be used to measure time for trajectories starting from a set of states}
\begin{equation}
t_2-t_1=1/\nu \frac{\sum_\alpha [W^L_{x_\alpha(t_2)}-W^L_{x_\alpha(t_1)}]A_{x_\alpha(t_1)}}{\sum_\alpha A_{x_\alpha(t_1)}}
\label{dtensstart}
\end{equation}
Multiplying Eq. \ref{wrndt2} by $A_i$ and summing over $i$ one finds that \textbf{the right eigenvector can be used to measure time for trajectories ending in a set of states}
\begin{equation}
t_2-t_1=1/\nu \frac{\sum_\alpha [W^R_{x_\alpha(t_2)}-W^R_{x_\alpha(t_1)}]A_{x_\alpha(t_2)}}{\sum_\alpha A_{x_\alpha(t_2)}}
\label{dtensend}
\end{equation}
For stationary processes, the averaging in Eqs. \ref{ensall}-\ref{dtensend} may include averaging over time, e.g., for Eq. \ref{dtensstart} one has
\begin{equation}
\Delta t=1/\nu \frac{\sum_{\alpha,t} [W^L_{x_\alpha(t+\Delta t)}-W^L_{x_\alpha(t)}]A_{x_\alpha(t)}}{\sum_{\alpha,t} A_{x_\alpha(t)}}.
\label{dtensstart2}
\end{equation}
\textbf{Time-reversed trajectories.} The equation for left eigenvector of time-reversed trajectories is
\begin{align*}
\sum_{j} n^T_{ji}(\Delta t)(W^{TL}_j-W^{TL}_i -\nu \Delta t)=0,
\end{align*}
which can be transformed to Eq. \ref{wrdef} with negative $\Delta t$
\begin{align*}
\sum_{j} n_{ij}(\Delta t)(W^{TL}_i-W^{TL}_j -\nu (-\Delta t))=0,
\end{align*}
 i.e., the right eigenvector for forward trajectories can be taken as the left eigenvector for time-reversed trajectories and vice-versa. 

\textbf{Time-dependent reaction coordinate.} By introducing $S^L_i(t)=W^L_i-\nu t$
and $S^R_i(t)=W^R_i-\nu t$ Eqs. \ref{wldef2} and \ref{wrdef} can be written as
\begin{align}
\sum_{j} n_{ji}[S^L_j(t+\Delta t)-S^L_i(t)]=0 \notag \\
\sum_{j} n_{ij}[S^R_i(t+\Delta t)-S^R_j(t)]=0
\label{sdef}
\end{align}
These equations can be considered as a generalization of the equation for the $p_{fold}$ reaction coordinate (Eq. \ref{rceq}) to time dependent reaction coordinates $S^L$ and $S^R$. For $\nu=0$, when the coordinates do not change with time and $n_{ij}=n_{ji}$ the single valued solutions equal $S^L=S^R=p_{fold}$. The equations, as well as Eq. \ref{rceq}, mean that the average change of the (time dependent) optimal coordinates along a trajectory is zero.

So far we have assumed that the optimal coordinate is a function of the state index $i$. Such a description is invariant with respect to the choice of the coordinate system. As shown in the illustrative examples below, it might be useful to embed the index in spatial coordinates, so that the optimal coordinate becomes a function of position $W(x)$. For example, in the one-dimensional case, one assigns position $x_i$ to state $i$ and assumes that  $W_{i+1}-W_i=\kbar \Delta x$, where $\kbar=k/2\pi=1/\lambda$ has the meaning of the wave number and $\lambda$ is the wavelength; the dimension of $k$ is inverse of $x$ to keep W dimensionless. In this case the change of the optimal coordinate can be written in the form where 
space and time are on an equal footing.
\begin{equation*}
S_{x+\Delta x}(t+\Delta t)-S_x(t)=\kbar \Delta x-\nu \Delta t
\end{equation*}


\textbf{Symmetric or relativistic coordinate.}
According to Eqs. \ref{dtensstart} and \ref{dtensend} one needs to use two different optimal coordinates $S^R$ and $S^L$ (or two additive eigenvectors) to describe incoming and outgoing or forward and time-reversed subsets of trajectories. It might be useful to introduce a single coordinate to describe all the subsets. The procedure is analogous to the simmetrization of the transition probability matrix $P_{ij}\rightarrow P_{ij}\sqrt{P^{st}_j/P^{st}_i}$ in the conventional case, which leads to the left and right eigenvectors being equal. Let
\begin{align*}
\sum_{j} P_{ji}(\Delta t)(S^L_j(t+\Delta t)-S^L_i(t))=0\\
\sum_{j} P_{ij}(\Delta t)P_j/P_i(S^R_i(t+\Delta t)-S^R_j(t))=0
\end{align*}
be two optimal coordinates that describe a stationary solution. Let $R_i=\sqrt{P_i}$, then
\begin{align*}
\sum_{j} P_{ji}(\Delta t)R_i/R_jR_j/R_i(S^L_j(t+\Delta t)-S^L_i(t))=&0\\
\sum_{j} P_{ij}(\Delta t)R_j/R_iR_j/R_i(S^R_i(t+\Delta t)-S^R_j(t))=&0.
\end{align*}
Introduce 
\begin{align}
S^L_j(t+\Delta t)-S^L_i&=R_i/R_j(S^s_j(t+\Delta t)-S^s_i(t)) \notag \\
S^R_j(t+\Delta t)-S^R_i&=R_j/R_i(S^s_j(t+\Delta t)-S^s_i(t)),
\label{slrs}
\end{align}
i.e., the change of $S^s$ is the geometric mean of the changes of $S^L$ and $S^R$.
Then
\begin{align*}
\sum_{j} P_{ji}(\Delta t)R_i/R_j(S^s_j(t+\Delta t)-S^s_i(t))&=0\\
\sum_{j} P_{ij}(\Delta t)R_j/R_i(S^s_i(t+\Delta t)-S^s_j(t))&=0
\end{align*}
or, if $\tilde{P}_{ij}$ is known,
\begin{align*}
\sum_{j} P_{ji}(\Delta t)R_i/R_j(S^s_j(t+\Delta t)-S^s_i(t))&=0\\
\sum_{j} \tilde{P}_{ij}(\Delta t)R_i/R_j(S^s_i(t+\Delta t)-S^s_j(t))&=0,
\end{align*}
or 
\begin{align*}
\sum_{j} \tilde{P}_{ji}(\Delta t)R_j/R_i(S^s_j(t+\Delta t)-S^s_i(t))&=0\\
\sum_{j} P_{ij}(\Delta t)R_j/R_i(S^s_i(t+\Delta t)-S^s_j(t))&=0.
\end{align*} 
Introducing $W^s_i-\nu^s t=S^s_i(t)$ one obtains
\begin{align}
\sum_{j} P_{ji}(\Delta t)R_i/R_j(W^s_j-W^s_i -\nu^s \Delta t)&=0\notag\\
\sum_{j} P_{ij}(\Delta t)R_j/R_i(W^s_i-W^s_j -\nu^s \Delta t)&=0,
\label{wsdef}
\end{align}
or
\begin{align}
\sum_{j} P_{ji}(\Delta t)R_i/R_j(W^s_j-W^s_i -\nu^s \Delta t)&=0\notag\\
\sum_{j} \tilde{P}_{ij}(\Delta t)R_i/R_j(W^s_i-W^s_j -\nu^s \Delta t)&=0,
\label{wsdef2}
\end{align}
or
\begin{align}
\sum_{j} \tilde{P}_{ji}(\Delta t)R_j/R_i(W^s_j-W^s_i -\nu^s \Delta t)&=0\notag\\
\sum_{j} P_{ij}(\Delta t)R_j/R_i(W^s_i-W^s_j -\nu^s \Delta t)&=0.
\label{wsdef3}
\end{align}$W^s$ is not an additive eigenvector, meaning that Eqs. \ref{slrs}-\ref{wsdef3} are valid only in the limit of $\Delta t\rightarrow 0$. They are
not valid for an arbitrarily large $\Delta t$, as we show later, since the symmetrized matrix is not a stochastic matrix. However, such a coordinate can be used in the limit of small $\Delta t$ to measure time for both starting and ending subsets of trajectories as
\begin{equation}
t_2-t_1= \frac{\sum_\alpha [W^s_{x_\alpha(t_2)}-W^s_{x_\alpha(t_1)}]A_{x_\alpha(t_1)}R^{-1}_{x_\alpha(t_2)}R^{-1}_{x_\alpha(t_1)}}{\sum_\alpha A_{x_\alpha(t_1)}R^{-1}_{x_\alpha(t_2)}R^{-1}_{x_\alpha(t_1)}}/\nu^s
\label{wsdtensstart}
\end{equation}
and
\begin{equation}
t_2-t_1=\frac{\sum_\alpha [W^s_{x_\alpha(t_2)}-W^s_{x_\alpha(t_1)}]A_{x_\alpha(t_2)}R^{-1}_{x_\alpha(t_2)}R^{-1}_{x_\alpha(t_1)}}{\sum_\alpha A_{x_\alpha(t_2)}R^{-1}_{x_\alpha(t_2)}R^{-1}_{x_\alpha(t_1)}}/\nu^s 
\label{wsdtensend}
\end{equation}
Eq. \ref{slrs} can be used to determine $S^s$ and $R$ from $S^L$ and $S^R$. 

\textbf{Equations for the rate matrix.} To derive the equations for the rate matrix we let $P_{ji}(\Delta t)=e^{\Delta t K_{ji}}\approx\delta_{ji} +\Delta t K_{ji}$, where $K_{ji}$ is the rate of going from state $i$ to state $j$ and $\sum_j K_{ji}=0$.
\begin{align}
\sum_{j} (\delta_{ji} + K_{ji} \Delta t) (W^L_j-W^L_i -\nu \Delta t) =0\notag \\
\sum_{j} K_{ji} \Delta t (W^L_j-W^L_i) -\nu \Delta t =0\notag \\
\sum_{j} K_{ji}(W^L_j-W^L_i) -\nu =0
\label{wlrate}
\end{align}
Similarly one obtains
\begin{align}
\sum_{j} \tilde{K}_{ij}(W^R_i-W^R_j) -\nu=0,
\label{wrrate}
\end{align}
where $\tilde{K}_{ij}=K_{ij}(\Delta t)P^{st}_j/P^{st}_i$. For the symmetric coordinate one obtains
\begin{multline*}
\sum_{j} (\delta_{ji} + K_{ji} \Delta t)R_i/R_j (W^s_j-W^s_i -\nu^s \Delta t) =0\\
\sum_{j} K_{ji} R_i/R_j (W^s_j-W^s_i) -\nu^s (1+\Delta t\sum_{j} K_{ji}R_i/R_j) =0\\
\end{multline*}
which shows that $W^s$ and $\nu^s$ become independent of $\Delta t$ in the limit of $\Delta t\rightarrow 0$, where Eq. \ref{wsdef} reads
\begin{align}
\sum_{j} K_{ji}R_i/R_j(W^s_j-W^s_i) -\nu^s &=0 \notag\\
\sum_{j} K_{ij}R_j/R_i(W^s_i-W^s_j) -\nu^s &=0,
\label{wsrate}
\end{align}
or, if $\tilde{K}_{ij}$ is known,
\begin{align}
\sum_{j} K_{ji}R_i/R_j(W^s_j-W^s_i) -\nu^s &=0 \notag\\
\sum_{j} \tilde{K}_{ij}R_i/R_j(W^s_i-W^s_j) -\nu^s &=0.
\label{wsrate2}
\end{align}

\subsection*{Illustrative Example 1.}
To illustrate the introduced concepts, consider the following example. Consider a system that moves to the right with rate $K_{i+1,i}=r_i$. For a small $\Delta t$ only $n_{i,i}$ and $n_{i+1,i}$ are non zero. For such a system the number of transitions from $i$ to $i+1$ is constant: $n_{i+1,i}=const=J\Delta t=r_i \Delta tn_i$, and $n_i=J/r_i$, where $J$ is flux. For the number of transitions from $i$ to $i$ one has: $n_{i,i}=(1-r_i \Delta t)n_i=J(1-r_i \Delta t)/r_i.$ For the left eigenvector optimal coordinate one finds (Eq. \ref{wldef2})
\begin{gather}
n_{i,i}(W^L_{i}-W^L_{i} -\nu \Delta t) +n_{i+1,i}(W^L_{i+1}-W^L_{i}-\nu \Delta t)=0\notag\\
-\nu +r_i(W^L_{i+1}-W^L_{i})=0\notag\\
W^L_{i+1}-W^L_{i}=\nu/r_i
\label{ex1wl}
\end{gather}
For the right eigenvector optimal coordinate one obtains (Eq. \ref{wrdef})
\begin{gather*}
n_{i,i}(W^R_{i}-W^R_{i} -\nu \Delta t) +n_{i,i-1}(W^R_{i}-W^R_{i-1}-\nu \Delta t)=0\\
-\nu/r_i +W^R_{i}-W^R_{i-1}=0\\
W^R_{i}-W^R_{i-1}=\nu/r_i\\
W^R_{i+1}-W^R_{i}=\nu/r_{i+1}
\end{gather*}
i.e., it is different from the left eigenvector.\\
The same result can be found using Eq. \ref{wrrate}.
$\tilde{K}_{i+1,i}=K_{i+1,i}P^{st}_i/P^{st}_{i+1}=r_i(J/r_i)/(J/r_{i+1})=r_{i+1}$. Hence,
\begin{gather*}
r_{i+1}(W^R_{i+1}-W^R_i)-\nu=0\\
W^R_{i+1}-W^R_i=\nu/r_{i+1}
\end{gather*}
For the symmetrized (relativistic) optimal coordinate one finds using Eqs. \ref{wsrate} ($R_i=\sqrt{J/r_i}$)
\begin{gather*}
r_iR_i/R_{i+1}(W^s_{i+1}-W^s_i)-\nu^s=0\\
r_iR_i/R_{i+1}(W^s_{i+1}-W^s_i)-\nu^s=0\\
\sqrt{r_ir_{i+1}}(W^s_{i+1}-W^s_i)-\nu^s=0\\
W^s_{i+1}-W^s_i=\nu^s/\sqrt{r_ir_{i+1}}
\end{gather*}
or, if one knows $\tilde{K}_{ji}$, then Eq. \ref{wsrate2} can be used to find both $R_i$ and $W^s_i$:
\begin{gather*}
r_iR_i/R_{i+1}(W^s_{i+1}-W^s_i)-\nu^s=0\\
r_{i+1}R_{i+1}/R_i(W^s_{i+1}-W^s_i)-\nu^s=0\\
r_iR_i/R_{i+1}=r_{i+1}R_{i+1}/R_i\\
R_i=\sqrt{J/r_i}\\
\sqrt{r_ir_{i+1}}(W^s_{i+1}-W^s_i)-\nu^s=0\\
W^s_{i+1}-W^s_i=\nu^s/\sqrt{r_ir_{i+1}}
\end{gather*}

Eq. \ref{ex1wl} defines the optimal coordinate as a function of index $i$. Index $i$ can be embedded into a spatial coordinate, i.e., each state ($i$) can be given a position $x_i$, so that the optimal coordinate is a function of the position. For example, if $x_i$ are selected as $x_{i+1}-x_i=c/r_i$, where $c$ is a constant with dimension of velocity, then $W^L=\kbar x$, $\nu=c\kbar$, and $S=W-\nu t=const$ describes a wave moving to the right with constant velocity of $c$. For the right coordinate one has
$x_{i+1}-x_i=c/r_{i+1}$, i.e., both coordinates can not be simultaneously embedded to keep $c$ constant. 

\begin{figure}[htbp]
\centering 
\resizebox*{\columnwidth}{!}{\includegraphics*[]{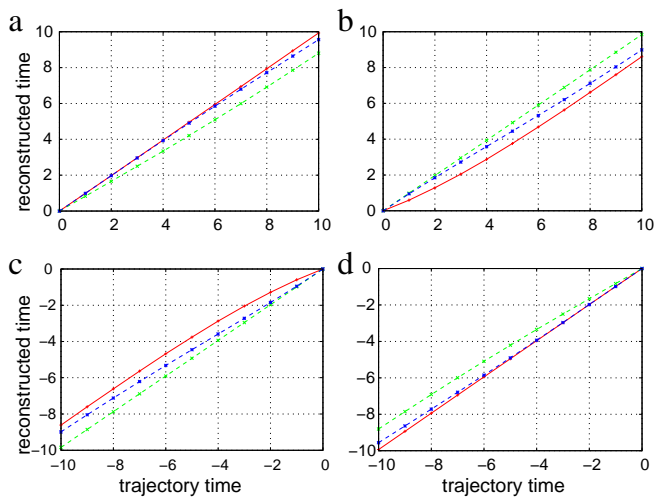}}
\caption{Reconstruction of time from system snapshots (configurations) of an ensemble of:
a) forward trajectories starting from x=50, b) forward trajectories ending at x=50, c) time-reversed trajectories starting from x=50, d) time-reversed trajectories ending at x=50. Reconstruction using $W^L$, $W^R$ and $W^s$ are shown by red, green and blue lines, respectively. The plots show the reconstructed time vs the actual time.}
\label{recontime}
\end{figure}

\textbf{Numerical example.}
Consider a system with $r_i=(mod(i,5)+5)/25$. 10000 trajectories have been simulated by Monte Carlo (with times step of $dt=0.1$) starting from $i=1$ until the system reached $i=100$. Fig. \ref{recontime} shows time reconstructed from the trajectories using left, right and symmetric optimal coordinates for $n\Delta t=1,...,10$ by applying corresponding variants of Eq. \ref{dtensstart2}.  Panel \textbf{a} shows that time reconstructed for trajectories starting from $i=50$, agrees with actual time if reconstruction is performed with the left coordinate and disagrees significantly if performed with the right coordinate. Panel \textbf{b} shows that time reconstructed for a subset of trajectories ending in $i=50$ is accurate if the right coordinate is used and not if the left one is used. The relativistic coordinate reconstructs time accurately for both sets of trajectories but only for relatively short time intervals. At longer time intervals  the reconstructed time deviates from the actual one. To accurately reconstruct time for long time intervals using the relativistic coordinate, the trajectory needs to be divided into short segments, time for which can be accurately reconstructed and then the total time can be found as their sum.  The left and right coordinates reconstruct times accurately for their corresponding sets of trajectories for arbitrary long trajectory segments. Panels \textbf{c} and \textbf{d} show results for time-reversed trajectories, where the left and right coordinates exchange their function.

\section*{Dynamics with detailed balance.} From now on we consider only systems with stationary dynamics where detailed balance holds $n_{ij}=n_{ji}$ and where, correspondingly, $\tilde{P}_{ji}(\Delta t)=P_{ij}(\Delta t)$ and $\tilde{K}_{ji}=K_{ij}$. For such systems, as can be easily seen, straightforward computation of right or left \textit{additive} eigenvectors leads to $\nu=0$. For example, by summing up over $i$  Eqs \ref{wldef2} one obtains
\begin{gather*}
\sum_{ij} n_{ij}(W^L_i-W^L_j -\nu \Delta t)=0\\  
\nu \Delta t\sum_{ij} n_{ij}=\sum_{ij} n_{ij}(W^L_i-W^L_j)=0  
\end{gather*}
Thus, solutions with nonzero $\nu$, necessary for the estimation of time intervals are not possible (in the space of single valued functions).

Solutions with $\nu\ne 0$ become possible, however, if one assumes that $W_i$ is not a single valued function, i.e., that the next time the system visits the same state $i$, $W_i$ can be different. One can suggest multiple reasons for that. For example, if a system moves on a line, it has to move in the reverse direction to return to the same point. The optimal coordinates that describe the motion in the backward  and forward directions should not necessary be the same. So each time the system changes direction, it may be described by a new coordinate. For systems moving on a ring the situation is more familiar. For example, for a random walk on a ring, considered above, the optimal coordinate equals $\phi$, the angular position (phase) on the ring, which covers the ring periodically. When the system returns to the same point by completing a cycle around the ring, the change in $W_i$ is analogous to the increase in the $\phi$ by $2\pi$. The classical action function is yet another example.

Note that the multivaluedness may lead to the following counter-intuitive property
\begin{align}
(W_i-W_j)+(W_j-W_i)\ne 0 \notag\\
(W_i-W_j)\ne -(W_j-W_i),
\label{nonzero}
\end{align}  
if $W_i$ in different brackets belong to different branches.

It seems that (to the best of my knowledge) the theory of such multivalued solutions for left and right additive eigenvectors has not been developed. I will present below some examples, where particular solutions can be found in a straightforward manner.
 
\subsection*{Reducing equations to a particular branch of a multivalued function}
While the equations on optimal coordinates are just simple systems of linear equations, they can not be solved with conventional linear algebra methods because the coordinates are multivalued functions. Assume that, for example, based on physical intuition, one knows where the transition between different branches of the multivalued function happens and that the difference between the branches is always the same (the solution is periodic). For example, if a new branch is reached at transition from $i$ to $j$ and the value at a new branch is related to the value at an old branch as $W^{new}_{j}=W_{j}+d_{ji}$, then  Eq. \ref{wldef} and Eq. \ref{wrdef2} can be rewritten for values at one (old) branch as
\begin{align}
\sum_{j} P_{ji}(\Delta t)(W^L_j+d_{ji} -W^L_i - \nu \Delta t)&=0
\label{wldefdij}\\
\sum_{j} \tilde{P}_{ij}(\Delta t)(W^R_i+d_{ij} -W^R_j-\nu \Delta t)&=0,
\label{wrdefdij}
\end{align}
where $d_{ij}$ are the differences (in phase) between different branches of the multivalued functions. The values at any branch can be taken since the solution is invariant to constant shift $W_i=W_i+c$. Assume further that any solution with many nonzero $d_{ij}$ can be represented as a linear combination of basis solutions with few or even single nonzero $d_{ij}$. Since the solution is defined up to a factor, for the latter case we can set the non-zero $d_{ij}=1$. 

For the rate matrix one obtains
\begin{align}
\sum_{j} K_{ji}(W^L_j+d_{ji} -W^L_i) - \nu&=0
\label{wlratedij}\\
\sum_{j} \tilde{K}_{ij}(W^R_i+d_{ij} -W^R_j)-\nu&=0,
\label{wrratedij}
\end{align}
For relativistic optimal coordinate, for example Eq. \ref{wsrate2} one obtains 
\begin{align}
\sum_{j} K_{ij}R_i/R_j(W^s_j+d_{ji} -W^s_i ) - \nu^s&=0\notag\\
\sum_{j} \tilde{K}_{ij}R_i/R_j(W^s_i+d_{ij} -W^s_j ) - \nu^s &=0.
\label{wrdij}
\end{align}

Alternatively, one can explicitly introduce multivaluedness by introducing variable $l$ that describes the current branch. The optimal coordinate becomes a function of two variables $W_{l,i}$, where one further assumes $W_{l,i}=ld+W_i$. For such defined optimal coordinates Eq. \ref{nonzero} is no longer counter-intuitive
\begin{align*}
(W_{l,i}-W_{l,j})+(W_{l+1,j}-W_{l,i})\ne 0 \\
(W_{l,i}-W_{l,j})+(W_{l,j}-W_{l,i})=0
\end{align*}  

\subsection*{Illustrative Example 2. Transitions between two states with different rates.}
Consider a system with dynamics described by the following master equation
\begin{align*}
\partial p_1/\partial t&=-r_1 p_1 +r_2 p_2\\
\partial p_2/\partial t&=-r_2 p_2 +r_1 p_1
\end{align*}
We assume that an optimal coordinate changes branches when the system makes transition $2\rightarrow 1$. The coordinate is taken in the form $W_{l,i}=l+W_i$.
For the left additive eigenvector one obtains (Eq. \ref{wlrate})
\begin{gather*}
r_1(W^L_{l,2}-W^L_{l,1}) -\nu=0\\
r_2(W^L_{l+1,1}-W^L_{l,2}) -\nu=0\\
[W^L_2-W^L_1]-\nu/r_1=0\\
[W^L_1+1-W^L_2]-\nu/r_2=0\\
\nu=1/(1/r_1+1/r_2)\\
W^L_1=0\\
W^L_2=\nu/r_1
\end{gather*}
Thus, one has $W^L_2-W^L_1=\nu/r_1$, $W^L_1-W^L_2=\nu/r_2$ and $(W^L_2-W^L_1)+(W^L_1-W^L_2)=1\ne 0$ because $W^L_1$ in the second bracket belongs to the next branch.\\
For stationary (equilbirum) populations one has
$P^{st}_1=1/r_1$ and $P^{st}_2=1/r_2$, and $\tilde{K}_{12}=K_{21}=r_1$,
$\tilde{K}_{21}=r_2$. For the right additive eigenvector one finds (Eq. \ref{wrrate}).
\begin{gather*}
r_2(W^R_{l,2}-W^R_{l,1}) -\nu=0\\
r_1(W^R_{l+1,1}-W^R_{l,2}) -\nu=0\\
\nu=1/(1/r_1+1/r_2)\\
W^R_1=0\\
W^R_2=\nu/r_2.
\end{gather*}
Thus while $\nu$ for both coordinates is the same, $W^L\ne W^R$.\\
\textbf{Explicit symmetrization}. For the symmetric rate matrix 
one obtains $K^s_{21}=K_{21}\sqrt{P^{st}_1/P^{st}_2}=r_1\sqrt{r_2/r_1}=\sqrt{r_1r_2}=K^s_{12}$. $W^s$ can be found as a left or right eigenvector of the symmetric rate matrix 
\begin{gather*}
\sqrt{r_1r_2}(W^s_{l,2}-W^s_{l,1}) -\nu^s=0\\
\sqrt{r_1r_2}(W^s_{l+1,1}-W^s_{l,2}) -\nu^s=0\\
\nu^s=\sqrt{r_1r_2}/2\\
W^s_1=0\\
W^s_2=1/2
\end{gather*}
\textbf{Implicit symmetrization}. The equation on the relativistic coordinate Eq. \ref{wsrate2} reads
\begin{align*}
r_1R_1/R_2(W^s_{l,2}-W^s_{l,1}) -\nu^s=0\\
r_2R_2/R_1(W^s_{l+1,1}-W^s_{l,2}) -\nu^s=0\\
r_2R_2/R_1(W^s_{l,2}-W^s_{l,1}) -\nu^s=0\\
r_1R_1/R_2(W^s_{l+1,1}-W^s_{l,2}) -\nu^s=0\\
\end{align*}
After substitution $W^s_{l,i}=l+W^s_i$ one finds both $W^s_i$ and $R_i$
\begin{gather*}
r_1R_1/R_2(W^s_2-W^s_1) -\nu^s=0\\
r_2R_2/R_1(W^s_1+1-W^s_2) -\nu^s=0\\
r_2R_2/R_1(W^s_2-W^s_1) -\nu^s=0\\
r_1R_1/R_2(W^s_1+1-W^s_2) -\nu^s=0\\
r_2R_2/R_1=r_1R_1/R_2\\
R_1=1/\sqrt{r_1}, \quad R_2=1/\sqrt{r_2}\\
\sqrt{r_1r_2}(W^s_2-W^s_1) -\nu^s=0\\
\sqrt{r_1r_2}(W^s_1+1-W^s_2) -\nu^s=0\\
\nu^s=\sqrt{r_1r_2}/2\\
W^s_1=0\\
W^s_2=1/2
\end{gather*}

\begin{figure}[htbp]
\centering 
\resizebox*{\columnwidth}{!}{\includegraphics*[]{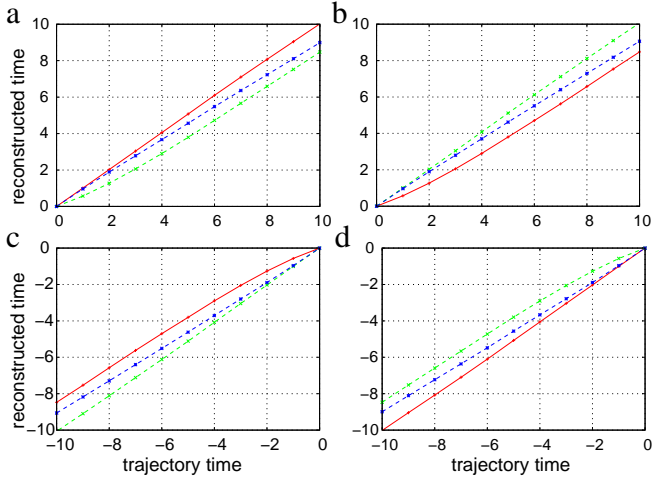}}
\caption{Reconstruction of time from system snapshots (configurations) from ensemble of:  a) forward trajectories starting from x=(10,1), b) forward trajectories ending at x=(10,1), c) time-reversed trajectories starting from x=(10,1), d) time-reversed trajectories ending at x=(10,1). Reconstruction using $W^L$, $W^R$ and $W^s$ are shown by red, green and blue lines, respectively.  The plots show the reconstructed time vs the actual time.}
\label{rtime_r1r2}
\end{figure}

\textbf{Numerical example.} 1000 trajectories (time series of $l,i$) each of length $10^5 dt$ were simulated by MC with time steps of $dt=0.01$ and saved with time interval of $\Delta t=1$. The transition rates are $r_1=0.1$ and $r_2=0.2$. Fig. \ref{rtime_r1r2} shows the reconstructed time vs the actual time. For forward trajectories, starting from a state, time can be reconstructed only by the left coordinate and conversely for forward trajectories ending in a state, time can be reconstructed only by the right coordinate. The relativistic coordinate can be used to reconstruct time in both cases but only for short time intervals.

\subsection*{Illustrative Example 3. Stochastic model of the telegraphers equation.}
Consider a particle that jumps in a constant direction, and changes direction with rate $r$. The model can be considered as a discrete version of the stochastic model of the telegraphers equation \cite{kac_stochastic_1974}. We assume that every time the direction is changed the dynamics is described by a new coordinate. Dynamics in the positive direction are described by coordinates $W_{2l+1,i}$, while that in the negative directions are described by coordinates $W_{2l+2,i}$. Thus we have the following set of transitions:\\
$W_{2l+1,i} \rightarrow$  to $W_{2l+1,i+1}$ with probability $1-r \Delta t$ or to $W_{2l+2,i-1}$ with probability $r \Delta t$. 
$W_{2l+2,i} \rightarrow$ to $W_{2l+2,i-1}$ with probability $1-r \Delta t$ or to $W_{2l+3,i+1}$ with probability $r \Delta t$.
For the left additive eigenvector one has (Eq. \ref{wldef})
\begin{align*}
(1-r \Delta t)[W^L_{2l+1,i+1}&-W^L_{2l+1,i}-\nu \Delta t]+\\&r \Delta t[W^L_{2l+2,i-1}-W^L_{2l+1,i} -\nu \Delta t]=0 \\   
(1-r \Delta t)[W^L_{2l+2,i-1}&-W^L_{2l+2,i}-\nu \Delta t]+\\&r \Delta t[W^L_{2l+3,i+1}-W^L_{2l+2,i} -\nu \Delta t]=0
\end{align*}
We assume that the (basis) solutions are periodic, i.e.,  $W^L_{l+m,i}=W^L_{l,i}+md$, in particular, we consider the case where $m=2$. Let $W^L_{2l+1,i}=2l+1+i \Delta x \kbar+w_1$, $W^L_{2l+2,i}=2l+2+(i+1) \Delta x \kbar+w_2$, i.e., index $i$ is considered to be embedded into coordinate $x$ as $i \Delta x\sim x $ and $W\sim \kbar x$. After substitution one finds    
\begin{align*}
(1-r \Delta t)\kbar\Delta x+r \Delta t [w_2+1-w_1] -\nu \Delta t=0\\     
-(1-r \Delta t)\kbar\Delta x+r \Delta t [w_1+1-w_2]-\nu \Delta t=0\\     
\nu=r
\end{align*}
We assume that $\Delta x=c \Delta t$, where $c$ is a constant, so that the limit $\Delta t \rightarrow 0$ exists. We let $w_1=0$ and find 
\begin{align*}
w_2&=-(1-r \Delta t) c\kbar/\nu\\
W^L_{2l+1,i}&=2l+1 + \kbar i \Delta x \\
W^L_{2l+2,i}&=2l+2 + \kbar (i+1) \Delta x   -(1-r \Delta t)c\kbar/\nu 
\end{align*}
In the limit $\Delta t\rightarrow0$ and, correspondingly, $\Delta x\rightarrow0$
\begin{align*}
W^L_{2l+1,x}&=2l+1 + \kbar x \\
W^L_{2l+2,x}&=2l+2 + \kbar x -c\kbar/\nu 
\end{align*}
For the right additive eigenvector 
\begin{align*}
(1-r \Delta t)[W^R_{2l+1,i}&-W^R_{2l+1,i-1}-\nu \Delta t]+\\&r \Delta t[W^R_{2l+1,i}-W^R_{2l,i-1} -\nu \Delta t]=0\\     
(1-r \Delta t)[W^R_{2l+2,i}&-W^R_{2l+2,i+1}-\nu \Delta t]+\\&r \Delta t[W^R_{2l+2,i}-W^R_{2l+1,i+1} -\nu \Delta t]=0\\
\end{align*}one analogously finds
\begin{gather*}
(1-r \Delta t)\kbar\Delta x+r \Delta t [w_1+1-w_2] -\nu \Delta t=0\\     
-(1-r \Delta t)\kbar\Delta x+r \Delta t [w_2+1-w_1]-\nu \Delta t=0\\     
\begin{align*}
\nu&=r\\
w_1&=0\\
w_2&=(1-r \Delta t) c\kbar/\nu\\
W^R_{2l+1,i}&=2l+1 + \kbar i \Delta x \\
W^R_{2l+2,i}&=2l+2 + \kbar (i+1) \Delta x   +(1-r \Delta t)c\kbar/\nu \\
W^R_{2l+1,x}&=2l+1 + \kbar x \\
W^R_{2l+2,x}&=2l+2 + \kbar x +c\kbar/\nu,
\end{align*}
\end{gather*}
i.e., $W^L\neq W^R$. Note, that equations for the left and right additive eigenvectors allow more complex solutions with quadratic dependence on x, but we do not consider them here. 

\textbf{The relativistic coordinate.} Since the left and right additive eigenvectors are different, it is useful to find the relativistic coordinate. Note, however, that the situation is slightly different from the one considered before. Here the transition matrix is symmetric and the left and right additive eigenvectors at $k=0$ (in the rest frame) are equal. They differ in a moving frame. We proceed analogously.

Let the left and right optimal coordinates be
\begin{align*}
\sum_{j} P_{ji}(\Delta t)(S^L_j(t+\Delta t)-S^L_i(t))&=0\\
\sum_{j} P_{ij}(\Delta t)(S^R_i(t+\Delta t)-S^R_j(t))&=0,
\end{align*}
where $P_{ij}=P_{ji}$. We introduce the symmetric (relativistic) reaction coordinate as
\begin{align}
S^L_j(t+\Delta t)-S^L_i(t)&=R_j/R_i(S^s_j(t+\Delta t)-S^s_i(t))\notag \\
S^R_i(t+\Delta t)-S^R_j(t)&=R_j/R_i(S^s_i(t+\Delta t)-S^s_j(t)).
\label{s2lrs}
\end{align}
Such a definition makes the comparison with the conventional relativistic equations of physics more straightforward. The equation is identical to Eq. \ref{slrs} if one makes substitution $R_i\rightarrow 1/R_i$ (or exchanges $S^L$ and $S^R$). One obtains 
\begin{align}
\sum_{j} P_{ji}(\Delta t)R_j/R_i(S^s_j(t+\Delta t)-S^s_i(t))&=0\notag \\
\sum_{j} P_{ij}(\Delta t)R_j/R_i(S^s_i(t+\Delta t)-S^s_j(t))&=0,
\label{s2def}
\end{align}
or
\begin{align}
\sum_{j} P_{ji}(\Delta t)R_j/R_i(W^s_j-W^s_i -\nu^s \Delta t)&=0 \notag \\
\sum_{j} P_{ij}(\Delta t)R_j/R_i(W^s_i-W^s_j -\nu^s \Delta t)&=0.
\label{ws2def}
\end{align}
The time intervals can be estimated as (substitute $R_i\rightarrow 1/R_i$ in Eqs. \ref{wsdtensstart} and \ref{wsdtensend})

\begin{equation}
t_2-t_1=\frac{\sum_\alpha [W^s_{x_\alpha(t_2)}-W^s_{x_\alpha(t_1)}]A_{x_\alpha(t_1)}R_{x_\alpha(t_2)}R_{x_\alpha(t_1)}}{\sum_\alpha A_{x_\alpha(t_1)}R_{x_\alpha(t_2)}R_{x_\alpha(t_1)}}/\nu^s 
\label{ws2dtensstart}
\end{equation}
and
\begin{equation}
t_2-t_1=\frac{\sum_\alpha [W^s_{x_\alpha(t_2)}-W^s_{x_\alpha(t_1)}]A_{x_\alpha(t_2)}R_{x_\alpha(t_2)}R_{x_\alpha(t_1)}}{\sum_\alpha A_{x_\alpha(t_2)}R_{x_\alpha(t_2)}R_{x_\alpha(t_1)}}/\nu^s 
\label{ws2dtensend}
\end{equation}
From Eq. \ref{ws2def}
\begin{align*}
(1-r \Delta t)&\frac{R_{2l+1,i+1}}{R_{2l+1,i}}[W^s_{2l+1,i+1}-W^s_{2l+1,i} -\nu^s \Delta t]+\\r \Delta t&\frac{R_{2l+2,i-1}}{R_{2l+1,i}}[W^s_{2l+2,i-1}-W^s_{2l+1,i} -\nu^s \Delta t]=0\\     
(1-r \Delta t)&\frac{R_{2l+2,i-1}}{R_{2l+2,i}}[W^s_{2l+2,i-1}-W^s_{2l+2,i} -\nu^s \Delta t]+\\r \Delta t&\frac{R_{2l+3,i+1}}{R_{2l+2,i}}[W^s_{2l+3,i+1}-W^s_{2l+2,i} -\nu^s \Delta t]=0\\     
(1-r \Delta t)&\frac{R_{2l+1,i-1}}{R_{2l+1,i}}[W^s_{2l+1,i}-W^s_{2l+1,i-1} -\nu^s \Delta t]+\\r \Delta t&\frac{R_{2l,i-1}}{R_{2l+1,i}}[W^s_{2l+1,i}-W^s_{2l,i-1} -\nu^s \Delta t]=0\\
(1-r \Delta t)&\frac{R_{2l+2,i+1}}{R_{2l+2,i}}[W^s_{2l+2,i}-W^s_{2l+2,i+1} -\nu^s \Delta t]+\\r \Delta t&\frac{R_{2l+1,i+1}}{R_{2l+2,i}}[W^s_{2l+2,i}-W^s_{2l+1,i+1} -\nu^s \Delta t]=0
\end{align*}
Assume that optimal coordinates are periodic for index $l$ with period 2, meaning  $R_{2l+1,i}=R_{1,i}$, $R_{2l+2,i}=R_{2,i}$. Since the system is translation invariant $R_{1,i}=R_1$, $R_{2,i}=R_2$. We assume, again, that $W^s_{2l+1,i}=2l+1+i \Delta x \kbar+w_1$, $W^s_{2l+2,i}=2l+2+(i+1) \Delta x \kbar+w_2$ and $\Delta x=c \Delta t$. After substitution and taking the limit $\Delta t \rightarrow 0$ (the equation for the relativistic coordinate is valid only in this limit)
\begin{align*}
\kbar c +r R_2/R_1(w_2+1-w_1)-\nu =0\\     
-\kbar c +r R_1/R_2(w_1+1-w_2) -\nu =0\\     
\kbar c +r R_2/R_1(w_1+1-w_2) -\nu=0\\
-\kbar c + r R_1/R_2(w_2+1-w_1)-\nu=0
\end{align*}
We dropped superscript s to simplify the notation.
By subtracting the third equation from the first, one finds that $w_2=w_1$, which we can set to 0, since the coordinate is defined up to a constant. Then one finds
\begin{gather}
\kbar c +r R_2/R_1 -\nu=0 \notag\\
-\kbar c + r R_1/R_2-\nu=0 \notag \\
\nu^2=r^2+c^2\kbar ^2 \notag \\
R_2/R_1=\sqrt{(\nu-c\kbar )/(\nu+c\kbar )} \notag \\
R_1=\sqrt{1+c\kbar/\nu}, \, \quad R_2=\sqrt{1-c\kbar/\nu},
\label{rel1}
\end{gather}
i.e., the infinite set of solutions, parametrized by $\kbar$ with relativistic relation between $\nu$ and $\kbar$, which is the reason 
behind naming the coordinate relativistic.

The stochastic dynamics projected on the optimal relativistic coordinate is described by $S=W-\nu t=const$, which  describes a plane wave running in (l,x) space with the phase velocity along $x$ of $\nu/\kbar$. To compute the group velocity, we consider a "wave packet" - two solutions with close but different values of $\kbar$ \cite{broglie}. Let their phases be equal at some point $\kbar_1 x +l-\nu_1 t=\kbar_2 x +l-\nu_2 t$. The equation for the phase agreement at the new position ($x+dx$) at next time instant ($t+dt$) is $\kbar_1(x+dx)+l-\nu_1(t+dt)=\kbar_2(x+dx)+l-\nu_2(t+dt)$. Hence $(\kbar_1-\kbar_2)dx=(\nu_1-\nu_2)dt$, or $v=dx/dt=(\nu_1-\nu_2)/(\kbar_1-\kbar_2)=\partial \nu/\partial \kbar=\kbar c^2/\nu$. Thus, one obtains $\nu=r/\sqrt{1-(v/c)^2}$, $\kbar=(vr/c^2)/\sqrt{1-(v/c)^2}$ and $R_1=\sqrt{1+v/c}$, $R_2=\sqrt{1-v/c}$. By introducing $E=h\nu$, $p=h\kbar$, $mc^2=hr$, where $h$ has the meaning of the Planck constant, one obtains the more familiar $E^2=m^2c^4+p^2c^2$, $v=pc^2/E$, $E=mc^2/\sqrt{1-(v/c)^2}$ and $p=mv/\sqrt{1-(v/c)^2}$.

Interpreting $R_i=\sqrt{P_i}$, where $P_i$ are the stationary probabilities one can compute the mean velocity $v=(cP_1-cP_2)/(P_1+P_2)=\kbar c^2/\nu$, which equals the group velocity. 


The relativistic coordinate can be found from the left and right additive eigenvectors using Eqs. \ref{s2lrs}. From 
\begin{align*}
\nu&=r\\
W^L_{2l+1,i}&=2l+1 + \kbar i \Delta x \\
W^L_{2l+2,i}&=2l+2 + \kbar (i+1) \Delta x   -(1-r \Delta t)c\kbar/\nu \\
W^R_{2l+1,i}&=2l+1 + \kbar i \Delta x \\
W^R_{2l+2,i}&=2l+2 + \kbar (i+1) \Delta x   +(1-r \Delta t)c\kbar/\nu
\end{align*}
one  computes
\begin{align*}
S^L_{2l+1,i+1}-S^L_{2l+1,i}=&\kbar \Delta x-\nu \Delta t=-(\nu-\kbar c)\Delta t\\
S^L_{2l+2,i-1}-S^L_{2l+1,i}=&1-(1-r \Delta t)c \kbar/\nu -\nu \Delta t=\\
&(1-r \Delta t)(1-c \kbar /\nu)\\
S^L_{2l,i-1}-S^L_{2l,i}=&
-(\nu+ \kbar c)\Delta t\\
S^L_{2l+1,i+1}-S^L_{2l,i}=&
(1-r \Delta t)(1+c \kbar /\nu)\\
S^R_{2l+1,i+1}-S^R_{2l+1,i}=&
-(\nu-\kbar c)\Delta t\\
S^R_{2l,i-1}-S^R_{2l,i}=&
-(\nu+\kbar c)\Delta t\\
S^R_{2l+2,i-1}-S^R_{2l+1,i}=&
(1-r \Delta t)(1+c \kbar /\nu)\\
S^R_{2l+1,i+1}-S^R_{2l,i}=&
(1-r \Delta t)(1-c \kbar /\nu),
\end{align*}
where we used shorthand notation for $S_j-S_i=S_j(t+\Delta t) -S_i(t)$. $P_{1,i}=P_1$, since $(S^L_{2l+1,i+1}-S^L_{2l+1,i})/(S^R_{2l+1,i+1}-S^R_{2l+1,i})=P_{1,i+1}/P_{1,i}=1$ and hence $S^s_{2l+1,i+1}-S^s_{2l+1,i}=S^L_{2l+1,i+1}-S^L_{2l+1,i}=-(\nu-\kbar c)\Delta t$, where $\nu=r$. Analogously $P_{2,i}=P_2$ and $S^s_{2l,i-1}-S^s_{2l,i}=S^L_{2l,i-1}-S^L_{2l,i}=-(\nu+\kbar c)\Delta t$. For transitions with the reversal of direction
\begin{align*}
P_2/P_1=(S^L_{2l+2,i-1}&-S^L_{2l+1,i})/(S^R_{2l+2,i-1}-S^R_{2l+1,i})=\\&(1-c\kbar/\nu)/(1+c \kbar/\nu)\\
S^s_{2l+2,i-1}-S^s_{2l+1,i}=&(1-r \Delta t)\sqrt{1-(c \kbar /\nu)^2}\\
S^s_{2l+1,i+1}-S^s_{2l,i}=&(1-r \Delta t)\sqrt{1-(c \kbar /\nu)^2}
\end{align*}
The obtained coordinate differs from the relativistic coordinate found before by an overall factor of $d=\sqrt{1-(c \kbar /\nu)^2}$, as can be seen by, e.g., computing $S^s_{2l+2,i}-S^s_{2l,i}$. By rescaling the coordinate $S^s\rightarrow S^s/d$, $\kbar\rightarrow\kbar/d$ and  $\nu\rightarrow \nu/d$ one finds that
\begin{gather*}
\nu=r/\sqrt{1-(c \kbar /\nu)^2}\\
r^2=\nu^2(1-(c \kbar /\nu)^2)=\nu^2-(c\kbar)^2
\end{gather*}

\begin{figure}[htbp]
\centering 
\resizebox*{\columnwidth}{!}{\includegraphics*[]{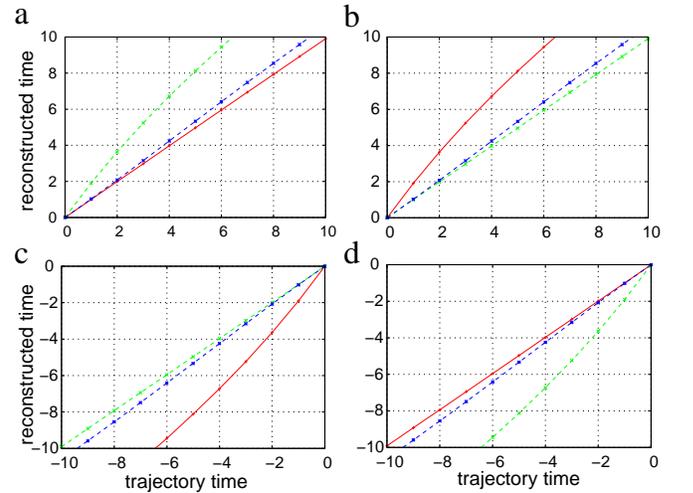}}
\caption{Reconstruction of time from system snapshots (configurations) from an ensemble of: a) forward trajectories starting from the positive direction, b) forward trajectories ending in the positive direction, c) time-reversed trajectories starting from the positive direction, d) time-reversed trajectories ending in the positive direction.  Reconstruction using $W^L$, $W^R$ and $W^s$ are shown by red, green and blue lines, respectively. The plots show the reconstructed time vs the actual time.}
\label{rtime_1de}
\end{figure}

\begin{figure}[htbp]
\centering 
\resizebox*{\columnwidth}{!}{\includegraphics*[]{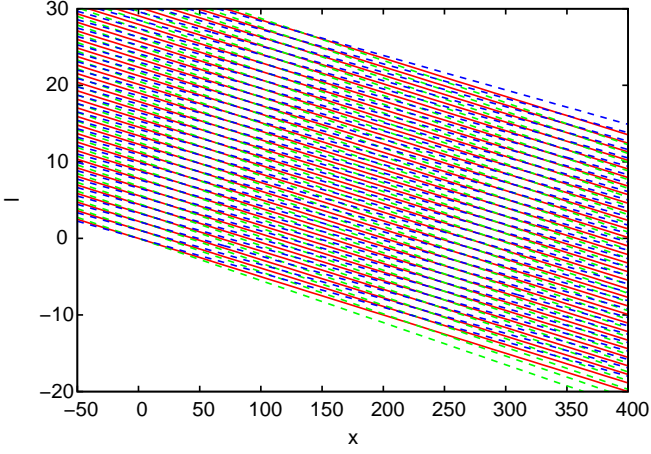}}
\caption{The dynamics of a wave packet. Surfaces of (relativistic) $S(l,x,t_i)=l+\kbar x -\nu t_i=const$ for $t_i=0,10,20 ... 300$ are shown for three solutions with $\kbar=\kbar_0$, $\kbar=1.1\kbar_0$ and $\kbar=0.9\kbar_0$ for $\kbar_0=0.05$. The region where phases are in agreement starts at (l=0,x=0,t=0) and moves with time along $x$ with the group velocity $v=\kbar/\nu$. }
\label{1dewavepacket}
\end{figure}

\textbf{Numerical example.}
1000 trajectories (time series of $l,i$) each of length $10^5 dt$ were simulated by MC with time steps of $dt=0.01$ and saved with time interval of $\Delta t=1$. The reversal rate is $r=0.1$. Fig. \ref{rtime_1de} shows times reconstructed with the optimal coordinates with $\kbar=0.05$. Relativistic coordinates with larger values of $\kbar$ correctly reconstruct time at shorter time intervals. Fig. \ref{1dewavepacket} shows the dynamics of a wave packet.

\subsection*{THE RELATION BETWEEN THE ADDITIVE AND CONVENTIONAL (MULTIPLICATIVE) EIGENVECTORS.}
An additive eigenvector is modified by matrix multiplication as $\mathbf{AW}^R=\mathbf{W}^R+\boldsymbol{\lambda}$, where $\boldsymbol{\lambda}=\{\lambda,...,\lambda\}$ is a vector where all components equals $\lambda$. An additive eigenvector is a multi-valued function of position, meaning that $j$ as a function of $W_j$ is a periodic function similar to $\exp(i2\pi W_j)$. A conventional eigenvector is modified by matrix multiplication as $\mathbf{A}\psi^R=\lambda\psi^R$. Eigenvectors of the master equation are often periodic functions. All this suggests that there might be a relation between an additive eigenvector and a phase (or logarithm) of a conventional eigenvector. Indeed, as we show below, under certain conditions it is possible to establish the correspondence. It, however, requires a certain modification of the acting operator, and correspondingly the underlying dynamics. The correspondence is similar to that between the classical action function and the wave function in quantum mechanics.

Let $\psi^R$, $\psi^L$ be the solutions of equations
\begin{align}
\psi_i^R(t+\Delta t)=\sum_j P_{ij}(\Delta t) \psi_j^R(t) \notag \\
\psi_i^L(t)=\sum_j P_{ji}(\Delta t) \psi_j^L(t+\Delta t),
\label{kolmeq}
\end{align}
where $\sum_j P_{ji}(\Delta t)=1$, or the corresponding continuous time equations.
If $\psi_i^L(t)=e^{i2\pi S^L_i(t)}=e^{i2\pi (W^L_i-\nu t)}$, where $\lambda=e^{-i2\pi \nu \Delta t}$ is the corresponding eigenvalue, then
\begin{align*}
\exp(i2\pi S^L_i(t))=\sum_j P_{ji}(\Delta t) \exp(i2\pi S^L_j(t+\Delta t))\\
1=\sum_j P_{ji}(\Delta t) \exp(i2\pi (S^L_j(t+\Delta t)-S^L_i(t)))
\end{align*}
Assume that $S^L_j(t+\Delta t)-S^L_i(t)$ are always close to 0 or some other integer number, i.e., 
$|S^L_j(t+\Delta t)-S^L_i(t) +d_{ji}|\ll 1$, where $d_{ji}$ is an integer, then one can expand the exponent and obtain Eq. \ref{wldefdij}.
\begin{align*}
1\approx \sum_j P_{ji}(\Delta t) [1+i2\pi (S^L_j(t+\Delta t)-S^L_i(t) +d_{ji})]\\
0\approx \sum_j P_{ji}(\Delta t) (S^L_j(t+\Delta t)-S^L_i(t) +d_{ji})\\
\sum_j P_{ji}(\Delta t) (W^L_j-W^L_i+d_{ji}-\nu \Delta t) \approx 0
\end{align*}

For the right eigenvector $\psi_j^R(t)=P_je^{i2\pi S^R_j(t)}$ one obtains Eq. \ref{wrdefdij}
\begin{align*}
P_i\exp(i2\pi S^R_i(t+\Delta t))=\sum_j P_{ij}(\Delta t)P_j \exp(i2\pi S^R_j(t))\\
1=\sum_j P_{ij}(\Delta t)P_j/P_i \exp(-i2\pi (S^R_i(t+\Delta t)-S^R_j(t)))\\
\sum_j P_{ij}(\Delta t)P_j/P_i (S^R_i(t+\Delta t)-S^R_j(t) -d_{ij}) \approx 0\\
\sum_j P_{ij}(\Delta t)P_j/P_i (W^R_i-W^R_j -d_{ij}-\nu \Delta t) \approx 0.
\end{align*}

Numerical analysis of the eigenvectors and eigenvalues of the system from illustrative example 1, shows that the relation is accurate for small eigenvalues and becomes inaccurate for large eigenvalues. To investigate the reason we consider the case $r_i=r$ analytically.
\begin{align*}
\exp(i2\pi (W_j-\nu (t+\Delta t))=&(1-r \Delta t)\exp(i2\pi (W_j-\nu t))\\&+r \Delta t\exp(i2\pi (W_{j-1}-\nu t))\\
\exp(-i2\pi\nu \Delta t)=(1-r \Delta t)&+r \Delta t\exp(i2\pi (W_{j-1}-W_j))\\
-i2\pi\nu=&r(\exp(i2\pi (W_{j-1}-W_j))-1)
\end{align*}
The equation reduces to $W_{j}-W_{j-1}=\nu/r$ only in the limit of $W_{j-1}-W_j\rightarrow0$ or $\nu\rightarrow 0$. One way to make it work for finite $\nu$ 
is to make the transition from $j-1$ to $j$ gradual by introducing intermediate  
states $j+k/n$, where $k=0,...,n-1$, so that corresponding $W_{j+(k+1)/n}-W_{j+k/n} \rightarrow 0$ for $n\rightarrow \infty$, while $W_j$ and $\nu$ stay the same. Let the time interval $\Delta t$ be further divided into $n$ sub-intervals. Instead of making a single jump from $j-1$ to $j$ with rate $r$ during $\Delta t$, the system makes n jumps from $j+(k-1)/n$ to $j+k/n$ with (yet unknown) rate $a$ each during $\Delta t/n$. The equations for the additive eigenvector are
\begin{align*}
a \Delta t/n (W_{j+(k+1)/n}-W_{j+k/n}) -\nu \Delta t /n=0.
\end{align*}
Summing the equations for $k=0,...,n-1$ one finds that $a=rn$.
The master equation is 
\begin{multline*}
p_{j+{k/n}}(t+\Delta t/n)=(1-rn \Delta t/n)p_{j+{k/n}}(t)+\\ rn \Delta t/n p_{j+{(k-1)/n}}(t)
\end{multline*} 
for $k=0,...,n-1$.
Let $x=\Delta x (j+k/n)$, then in the limit $n\rightarrow \infty$ one can approximate the finite differences by derivatives and obtain:
\begin{gather}
p_{j+{k/n}}+\Delta t/n \partial p_{j+{k/n}}/\partial t =(1-r \Delta t)p_{j+{k/n}}\notag+\\r \Delta t(p_{j+{k/n}}-1/n\partial p_{j+{k/n}}/\partial j) \notag\\
\Delta t/n \partial p_{j+{k/n}}/\partial t +r \Delta t/n\partial p_{j+{k/n}}/\partial j=0 \notag\\
\partial p_{j+{k/n}}/\partial t +r \partial p_{j+{k/n}}/\partial j=0 \notag\\
\partial p(x,t)/\partial t+c\partial p(x,t)/\partial x=0,
\label{vop1}
\end{gather}
where $c=r\Delta x$. The eigenfunction of the equation is $\exp(i2\pi(\nu x/c+\nu t))=\exp(i2\pi(W_x-\nu t))$. Thus, we have found the equation with (multiplicative) eigenfunction and eigenvalue, which correspond exactly to \textit{additive} eigenvector and eigenvalue. However, in order to do that it was necessary to modify the underlying dynamics of the system. First, the dynamics is nor longer stochastic. The differential operator describes a deterministic running wave. Second, the configuration space of the system has been extended. Instead of being integer $j\in Z$ it became real $x\in R$. It seems reasonable to name operators such as in Eq. \ref{vop1} virtual operators, since they describe virtual dynamics, not the actual dynamics of the system and are just a mathematical tool to obtain (multiplicative) eigenfunction and eigenvalue, which correspond exactly to \textit{additive} ones.

For the relativistic coordinate the correspondence is established analogously.
Let $P_{ij}(\Delta t)$ be such that the solutions of Eqs. \ref{kolmeq}
can be expressed as
\begin{align*}
\psi^R_j=\psi^L_j=R_je^{i2\pi S^s_j(t)}=R_je^{i2\pi (W^s_j-\nu t)}
\end{align*}
Then
\begin{gather*}
e^{-i2\pi \nu^s\Delta t}R_ie^{i2\pi W^s_i}=\sum_j P_{ij}(\Delta t)R_je^{i2\pi W^s_j}\\
e^{-i2\pi \nu^s\Delta t}=\sum_j P_{ij}(\Delta t)R_j/R_ie^{-i2\pi (W^s_i-W^s_j)}\\
1-i2\pi \nu^s\Delta t\approx \\\sum_j (\delta_{ij}+\Delta t K_{ij})R_j/R_i[1-i2\pi (W^s_i-W^s_j +d_{ij})],
\end{gather*}
where imaginary part equals
\begin{align*}
\sum_j K_{ij}R_j/R_i(W^s_i-W^s_j+d_{ij})-\nu^s \approx 0.
\end{align*}
For the left eigenvector one obtains
\begin{align*}
\sum_j K_{ji}R_j/R_i(W^s_j-W^s_i+d_{ji})-\nu\approx0.
\end{align*}
The two equations are the multi-valued rate matrix versions of Eq. \ref{ws2def}.

Consider system with dynamics described by the following master equation
\begin{align*}
p_1(t+ \Delta t)=(1-r \Delta t) p_1 +r \Delta t p_2\\
p_2(t+\Delta t)=(1-r \Delta t)p_2 +r \Delta t p_1
\end{align*}
where $p_1$ and $p_2$ are the probabilities to be in state 1 and 2, respectively, which is equivalent to the system considered in illustrate example 2 if one sets $r_1=r_2=r$. For this system, the left, right and relativistic coordinates are the same $W_{l,1}=l$, $W_{l,2}=l+1/2$ and $\nu=r/2$. 

The equation has two eigenvalues $\lambda=1$ and $\lambda=1-2r \Delta t$, which correspond to ($\lambda=\exp(i2\pi\nu \Delta t )$) 
$\nu=0$ and $\nu=ir/\pi$. The eigenvector of the second eigenvalue is $\psi_1=1=e^{-2\pi i 0}=e^{-2\pi i W_1}$ and $\psi_2=-1=e^{-2\pi i 1/2}=e^{-2\pi i W_2}$, in agreement with the additive eigenvectors.

The second eigenvalue is not in correspondence because $W_1-W_2$ is not small and the exponent can not be expanded just to linear terms. Since after two steps the systems  returns to itself, each step corresponds to rotation on $\pi$ radians. To make the linear exponent expansion accurate, for the correspondence to be valid, each step should be made infinitesimally small. Analogous to the above, one way to do this, is to make the rotation gradual, i.e., instead of rotation on $\pi$ radians with rate $r$, make $n$ rotations on $\pi/n$ radians with rate $nr$ where $n\rightarrow \infty$. Let time interval $\Delta t$ be further divided into $n$ sub-intervals and let $p_{1+j/n}$ represent the intermediate values, representing rotation by angle of $\pi/n$.
The equation for the additive eigenvector are 
\begin{align*}
rn \Delta t/n (W_{1+(j+1)/n}-W_{1+j/n}) -\nu \Delta t /n=0
\end{align*}
for $j=1,...,2n$. 
Master equation is
\begin{multline*}
p_{1+j/n}(t+\Delta t/n)=(1 -rn \Delta t/n) p_{1+j/n}(t) +\\rn \Delta t/n p_{1+(j-1)/n}(t).
\end{multline*}
If $n$ is large, one can expand the finite-difference equation and obtain
\begin{gather*}
\Delta t/n \partial p_{1+j/n}/\partial t=-rn \Delta t/n\partial p_{1+j/n}/\partial j\\
\partial p_{\phi}/\partial t=-r\pi\partial p_{\phi}/\partial \phi,
\end{gather*}
where $\phi=2\pi j/2n$ is the rotation angle. The equation has 
eigenfunction $p=e^{i\phi -ir\pi t}$ with eigenvalue $\mu=ir\pi$ corresponding to $\nu=r/2$. The eigenfunction at points $\phi=0$ and $\phi= \pi$ corresponds to $W_1$ and $W_2$. Thus, in order to obtain the correct correspondence between the
additive and multiplicative eigenvector and eigenvalue the stochastic process
had to be modified. The new process consists of infinitesimal jumps instead of finite jumps and it describes deterministic rotation instead of the original stochastic dynamics. The new process suggests that the system can have any $\phi$, while in the original process only $\phi=0$ and $\phi=\pi$ are possible.

In the previous construction many intermediate $p_{1+j/n}$ were introduced to explicitly represent the rotation by a small angle of $\pi/n$. The rotation can be also represented by a rotation matrix in some ($e_x$, $e_y$) basis. As $e_x$ and $e_y$ one can take unit vectors associated with $p_1$ and $p_{1+1/2}$. $p_2$ corresponds to $-e_x$ and can not be taken as basis vector because a rotation can not be represented as a linear sum of $e_x$ and $-e_x$. Each $p_{1+j/n}=xe_x+ye_y$ is a linear combination of the basis vectors with coefficients $x,y$. Since $x$ and $y$ are coordinates and not probabilities, they can be negative. When the system makes transition from $p_{1+(j-1)/n}$ to $p_{1+j/n}$, $x,y$ coordinates are changed by the rotation matrix
\begin{equation*}
\begin{pmatrix}
 \cos \pi/n &  -\sin \pi/n  \\
 \sin \pi/n & \cos \pi/n
\end{pmatrix}.
\end{equation*}
So the master equation is 

\begin{multline*}
\begin{matrix}
x(t+\Delta t/n)\\y(t+\Delta t/n)
\end{matrix}=
(1-rn \Delta t/n)
\begin{matrix}
x(t)\\y(t)
\end{matrix}
+\\rn \Delta t/n \begin{pmatrix}
 \cos \pi/n &  \sin \pi/n  \\
 -\sin \pi/n & \cos \pi/n
\end{pmatrix}  
\begin{matrix}
x(t)\\y(t)
\end{matrix},
\end{multline*}
the rotation matrix for angle $-\pi/n$ is taken to express $p_{1+(j-1)/n}$ from $p_{1+j/n}$.
Expanding the equation one obtains
\begin{align*}
x+\Delta t/n \partial x/\partial t=& x - rn \Delta t/n x + rn \Delta t/n x + rn\pi \Delta t/n^2 y\\
y+\Delta t/n \partial y/\partial t=& y - rn \Delta t/n y + rn \Delta t/n y - rn\pi \Delta t/n^2 x\\
\partial x/\partial t=&  r\pi y\\
\partial y/\partial t=& - r\pi x.
\end{align*}
The equation is similar to the one-dimensional relativistic Dirac equation for an electron in its rest frame. However, the original system operates only with $p_1$ and $p_2$; $p_{1+1/2}$ can not be observed. To alleviate this, the original cycle ($1\rightarrow 2\rightarrow 1$) can be extended to cycle $1\rightarrow 2 \rightarrow 3 \rightarrow 4 \rightarrow 1$, where state 3 is identical to 1 and 4 to 2. The additive eigenvalue and eigenvector are $\nu=r/4$, and $W_{l,1}=l$, $W_{l,2}=l+1/4$, $W_{l,3}=l+1/2$ and $W_{l,4}=l+3/4$. In this system $e_1$ and $e_2$ are associated with $p_1$ and $p_2$, and the virtual operator is (the rotation rate now is $rn\cdot \pi/2/n=r\pi/2$)
\begin{align}
da_1/dt & = r\pi/2 a_{2}\notag\\
da_{2}/dt  & = -r\pi/2 a_{1}\notag \\
a_3 & = -a_1\notag \\
a_4 & = -a_2
\label{vop2}
\end{align}
The eigenvalue $\mu=i\pi r/2$ corresponds to the additive eigenvalue of $\nu=r/4$, and eigenfunction is $\psi_1=1/2=e^{0}/2$, $\psi_2=i/2=e^{i\pi/2}/2$, $\psi_3=-1/2$ and $\psi_4=-i/2$, which is in correspondence with the additive eigenvector.

The dynamics described by the stochastic telegraphers equations is the superposition of constant motion to the left or to the right and change with rate $r$ between the two motions (directions). Hence the virtual operator for this equation is the superposition of Eqs. \ref{vop1} and \ref{vop2}
\begin{align*}
\partial a_1/\partial t + c\partial a_1 /\partial x&=r\pi/2 a_2 \notag \\
\partial a_2/\partial t - c\partial a_2 /\partial x&=-r\pi/2 a_1 \notag\\
a_3&=-a_1\notag\\
a_4&=-a_2
\end{align*}
which is equivalent to the one dimensional Dirac equation, if one denotes
$r\pi/2$ as $mc^2/\hbar$ or $mc^2=hr/4$.
The eigenvector of the virtual operator
\begin{align*}
a_1&=\sqrt{1+c\kbar/\nu}e^{i2\pi(\kbar x -\nu t)}\\
a_2&=\sqrt{1-c\kbar/\nu}e^{i2\pi(\kbar x -\nu t+1/4)}\\
a_3&=\sqrt{1+c\kbar/\nu}e^{i2\pi(\kbar x -\nu t+2/4)}\\
a_4&=\sqrt{1-c\kbar/\nu}e^{i2\pi(\kbar x -\nu t+3/4)},
\end{align*}
where $\nu^2=(r/4)^2+\kbar^2c^2$ is in agreement with the optimal coordinate (Eq. \ref{rel1}). The factor of 4, compare to the solution given by Eq. \ref{rel1}, is due to different normalization of optimal coordinates $W_{l+4,i}=W_{l,i}+1$ vs $W_{l+2,i}=W_{l,i}+2$. 

Note that two reversals of the direction $1\rightarrow2\rightarrow3$ (which result in the original direction) lead to the change of sign $a_1\rightarrow a_3=-a_1$. It requires four reversals of the direction to return to the original sign, analogous to the transformation of a spinor under $2\pi$ or $4\pi$ rotation. 

Thus, the change of direction during a random walk can be transformed to the virtual continuous operator representing rotation (in internal space). Every equilibrium stochastic dynamics, by definition, contains movements in opposite directions, meaning that virtual operators representing rotations are ubiquitous. 

The strategy of finding the virtual operator can be summarized as follows. The correspondence holds if $P(\Delta t)$ is such that $W_i-W_j\approx 0$ and $\nu$ is real. If it is not the case, then the configuration space is expanded with intermediate states (denoted by fractional index $i+k/n$) on which virtual dynamics described by a virtual operator $A(\Delta t/n)$ is introduced, such that $W_{i+(k+1)/n}-W_{i+k/n}\rightarrow 0$ with $n\rightarrow \infty$, while $W_{i}$ and $\nu$ do not change. For such an operator the correspondence between the additive and multiplicative eigenvectors and eigenvalues is exact. Hence, the correspondence is exact between the additive eigenvectors and eigenvalues of the original $P$ and the conventional 
eigenvectors and eigenvalues of the virtual operator $A$ on the original configuration space (integer index).

\subsection*{Illustrative Example 4. The telegraphers equation in a slowly varying potential.}
Let the reversal rate now be a function of the position ($r_i$), corresponding to a random walk in a potential.
\begin{align*}
(1-r_i \Delta t)\frac{R_{2l+1,i+1}}{R_{2l+1,i}}&[W^s_{2l+1,i+1}-W^s_{2l+1,i} -\nu^s \Delta t]+\\r_i \Delta t\frac{R_{2l+2,i-1}}{R_{2l+1,i}}&[W^s_{2l+2,i-1}-W^s_{2l+1,i} -\nu^s \Delta t]=0\\     
(1-r_i \Delta t)\frac{R_{2l+2,i-1}}{R_{2l+2,i}}&[W^s_{2l+2,i-1}-W^s_{2l+2,i} -\nu^s \Delta t]+\\r_i \Delta t\frac{R_{2l+3,i+1}}{R_{2l+2,i}}&[W^s_{2l+3,i+1}-W^s_{2l+2,i} -\nu^s \Delta t]=0\\     
(1-r_i \Delta t)\frac{R_{2l+1,i-1}}{R_{2l+1,i}}&[W^s_{2l+1,i}-W^s_{2l+1,i-1} -\nu^s \Delta t]+\\r_i \Delta t\frac{R_{2l,i-1}}{R_{2l+1,i}}&[W^s_{2l+1,i}-W^s_{2l,i-1} -\nu^s \Delta t]=0\\
(1-r_i \Delta t)\frac{R_{2l+2,i+1}}{R_{2l+2,i}}&[W^s_{2l+2,i}-W^s_{2l+2,i+1} -\nu^s \Delta t]+\\r_i \Delta t\frac{R_{2l+1,i+1}}{R_{2l+2,i}}&[W^s_{2l+2,i}-W^s_{2l+1,i+1} -\nu^s \Delta t]=0
\end{align*}
Let  $R_{2l+1,i}=R_{1,i}$, $R_{2l+2,i}=R_{2,i}$, $W^s_{l,i+1}-W^s_{l,i}=\kbar_i\Delta x$, $W^s_{2l+3,i}=2+W^s_{2l+1,i}$ and $W^s_{2l+2,i}=2+W^s_{2l,i}$. Assume that $r_i$ changes slowly with $i$ (fine discretization), meaning $\kbar_i\approx \kbar_{i+1}$ and $R_{1,i+1}/R_{1,i}\approx R_{2,i+1}/R_{2,i} \approx 1$, then one arrives at (we dropped superscript s)
\begin{align*}
\kbar_ic+r_i\frac{R_{2,i-1}}{R_{1,i}}[W_{2l+2,i-1}-W_{2l+1,i}] -\nu=0\\     
-\kbar_ic+r_i\frac{R_{1,i+1}}{R_{2,i}}[W_{2l+3,i+1}-W_{2l+2,i}] -\nu=0\\     
\kbar_ic+r_i\frac{R_{2,i-1}}{R_{1,i}}[W_{2l+1,i}-W_{2l,i-1}] -\nu=0\\
-\kbar_ic+r_i\frac{R_{1,i+1}}{R_{2,i}}[W_{2l+2,i}-W_{2l+1,i+1}] -\nu=0
\end{align*}
\begin{align*}    
\kbar_ic+r_i\frac{R_{2,i-1}}{R_{1,i}} -\nu=0\\     
-\kbar_ic+r_i\frac{R_{1,i+1}}{R_{2,i}} -\nu=0\\     
r^2_i=\nu^2-(\kbar_ic)^2
\end{align*}
Since $\kbar_i$ changes slowly with $i$ one can use continuous representation, where $\kbar(x)=\partial W(x)/ \partial x $ and the last equation becomes
\begin{equation*}
r^2(x)+c^2(\partial W(x)/ \partial x)^2 -\nu^2=0.
\end{equation*}
 Or for $S=W(x)-\nu t$
\begin{equation*}
r^2(x)+c^2(\partial S/ \partial x)^2 -(\partial S/ \partial t)^2=0
\end{equation*}
the (dimensionless) relativistic Hamilton-Jacobi equation with mass that is a function of coordinate.

In the derivation it was, again, assumed that $\Delta x =c \Delta t$, which can be considered as the property of the stochastic model. If one, however, assumes that the stochastic model is a microscopic model of (one-dimensional) general relativity, then
the speed of light is the universal constant only in local inertial frames of reference. For small velocities, i.e., small potential $U_i\ll r$, where $r_i=r+U_i$, the relativistic effects are negligible. In this case $\nu=r+e$, where $e\ll r$, and one obtains for $S=W(x)-et$ $$c^2(\partial S(x)/ \partial x)^2/2r +U(x) +\partial S/ \partial t=0$$ the classical (dimensionless) Hamilton-Jacobi equation. The dimensionality can be restored by multiplying $W$ and $S$ by $h$ and replacing $h\nu=E$, $hr=mc^2$, $\kbar h =p$.

\subsection*{Illustrative Example 5. Random walk with rate r.}
Consider a random walk on the line, where a system jumps to the nearby left or right state with rate r. Coordinate $W^s_{2l+1,i}$ describes movement to the right or when the system stays in the same state, and $W^s_{2l+2,i}$ describes movement to the left or when the system stays in the same state. In other words the optimal coordinate changes together with the direction. Equations on the relativistic coordinate are (superscript s is omitted)

\begin{align*}
r \Delta t\frac{R_{2l+1,i+1}}{R_{2l+1,i}}&[W_{2l+1,i+1}-W_{2l+1,i}-\nu \Delta t]+(1-2r \Delta t)\times \\ [ - \nu \Delta t]+r\Delta &t\frac{R_{2l+2,i-1}}{R_{2l+1,i}}[W_{2l+2,i-1}-W_{2l+1,i}-\nu \Delta t] =0\\     
r\Delta t\frac{R_{2l,i-1}}{R_{2l,i}}&[W_{2l,i-1}-W_{2l,i}-\nu \Delta t]+(1-2r \Delta t)\times \\ [ -\nu \Delta t]+r\Delta &t\frac{R_{2l+1,i+1}}{R_{2l,i}}[W_{2l+1,i+1}-W_{2l,i}-\nu \Delta t] =0\\     
r\Delta t\frac{R_{2l+1,i-1}}{R_{2l+1,i}}&[W_{2l+1,i}-W_{2l+1,i-1}-\nu \Delta t]+(1-2r \Delta t)\times \\ [-\nu \Delta t]+r\Delta &t\frac{R_{2l,i-1}}{R_{2l+1,i}}[W_{2l+1,i}-W_{2l,i-1}-\nu \Delta t] =0\\
r\Delta t\frac{R_{2l+2,i+1}}{R_{2l+2,i}}&[W_{2l+2,i}-W_{2l+2,i+1}-\nu \Delta t]+(1-2r \Delta t)\times\\ [-\nu \Delta t]+r\Delta &t\frac{R_{2l+1,i+1}}{R_{2l+2,i}}[W_{2l+2,i}-W_{2l+1,i+1}-\nu \Delta t]  =0
\end{align*}
Analogous with the above we assume $R_{2l+1,i}=R_{1}$, $R_{2l+2,i}=R_{2}$ and $W_{2l+1,i}=w_1+i\kbar \Delta x +2l+1$ and $W_{2l+2,i}=w_2+(i+1)\kbar \Delta x+2l+2$. 
\begin{align*}
r\kbar \Delta x +rR_2/R_1[w_2+1-w_1] -\nu=0\\     
-r\kbar \Delta x +rR_1/R_2[w_1+1-w_2] -\nu=0\\     
r\kbar \Delta x +rR_2/R_1[w_1+1-w_2] -\nu=0\\
-r\kbar \Delta x +rR_1/R_2[w_2+1-w_1]-\nu=0
\end{align*}
Solving, one finds $(\nu/r)^2=1+(\kbar \Delta x)^2$, i.e., the relativistic spectrum of a particle with mass 1, where $r$ and $\Delta x$ define the temporal and spatial scales. Or, analogous to the above, $\nu^2=r^2+\kbar^2 c^2$ if one denotes $c=r \Delta x$. Thus, the obtained results are not a peculiarity of the telegraphers model.

\section*{Discussion}
The problem of determining an optimal coordinate that describes dynamics in general has been considered. It has been shown that the problem is closely related to 
the problem of reconstructing time from a trajectory and the problem of defining the eigen-modes for stochastic dynamics. They are solved by introducing \textit{additive} eigenvectors. The eigenvectors are modified under the action of a stochastic matrix in a simple way $\mathbf{W^LP}=\mathbf{W^L}+\boldsymbol{\lambda}$. Such left and right additive eigenvectors can be used to reconstruct time from ensembles of trajectories starting or ending in a set of states, respectively. The symmetric or relativistic coordinate can be introduced. It allows one to reconstruct time for both ensembles of trajectories, but only for relatively small time intervals. For the dynamics with detailed balance the additive eigenvectors are multi-valued functions. It was shown that it is possible to establish a correspondence between an additive eigenvector and an eigenvalue and a conventional eigenvector and an eigenvalue of a virtual operator. The virtual operator, however, describes different dynamics in an extended configuration space. In particular, the virtual operator for a random walk on the line corresponds to the one-dimensional Dirac equation. 

The close relation between the equations describing stochastic dynamics and that of quantum mechanics is well known \cite{nelson_derivation_1966, gaveau_relativistic_1984}. In particular, analytical continuation, e.g., $t\rightarrow it$,  is a straightforward way to obtain the Schr\"{o}dinger equation from the diffusion equation or the one-dimensional Dirac equation from the telegraphers equation \cite{gaveau_relativistic_1984}. The presented results differ in the following. First, no analytic continuation is performed. Second, the resulting Dirac equation is a virtual operator, i.e., it does not describe the actual dynamics, it is just a mathematical tool to match the eigenvectors and eigenvalues. Third, the results are valid for generic one-dimensional random walks, no specific stochastic process is selected. 
Interestingly, the $l$ coordinate that explicitly keeps track of the branches of the multi-valued functions (or rather its continuous analog) seems analogous to the action coordinate in the 5 optics of Rumer \cite{rumer_1956}. In the 5 optics all physical quantities are periodic along the action coordinate and its period equals the Planck constant (or 1 in dimensionless units). 

Equations with detailed balance have no additive eigenvectors with $\nu\ne 0$ in the space of single valued functions. In order to obtain solutions with $\nu\ne 0$ we postulate that additive eigenvectors are multi-valued functions, i.e., we have enlarged the configuration space of the solutions. In particular, by introducing an additional variable which explicitly describes the branches of the multi-valued function. The whole construction may seem artificial at first. However, it could be viewed as being analogous to the introduction of complex numbers. Complex numbers have real and imaginary parts and are necessary to describe all the solutions of a polynomial equation just with real coefficients. As illustrated above, $W_i$ and $R_i$ can be considered as the polar representation of a complex number.

The purpose of an optimal coordinate being multi-valued and the difference from the conventional description can be illustrated as follows. 
Consider a system that stochastically transits between two states 1 and 2 (illustrative example 2). Let an ensemble of such systems be initially in state 1. With time some systems will transit to state 2 and then some of them will return to state 1. State 1 now contains two sets of systems: the systems which came back there from state 2 and the systems which never left state 1. The future dynamics of the two sets are described by the same set of equations and, conventionally, one considers them identical and count them together. However after such mixing, the information about the past dynamics (which was different) is lost, one can not reconstruct dynamics back in time. The multi-valuedness (of an optimal coordinate) is used to distinguish the two sets. The systems which came back from state 2 now belong to a different branch and thus the two sets can be distinguished. 

The branches of the optimal coordinate can be straightforwardly computed from the system trajectory if it is known with sufficiently fine temporal resolution. That is how it was done in the numerical examples and how it can be done in a real-life experiment.
If an experimental system does not allow the observation of a trajectory with sufficient temporal resolution, then, in principle, one may attempt to infer 
the branches from auxiliary variables. For example, the dynamics of a molecular motor or an enzyme might be described by an optimal coordinate with ring topology. The auxiliary variables for such systems could be the position of the motor along a track or the number of ATP/substrate/product molecules.

The transition to the next branch of an optimal coordinate when a system returns to a state visited before can be compared to the phenomenon of the geometric phase, i.e., the increment of the phase acquired when a quantum mechanical \cite{berry_anticipations_1990} or stochastic system \cite{astumian_adiabatic_2007, sinitsyn_universal_2007, rahav_directed_2008} is undergoing adiabatic cyclic evolution in parameter space. In the case of stochastic systems one considers the dynamics to be described by the master equation with detailed balance. The equilibrium net flux between any two states is therefore zero. If the parameters (rates) of the master equation are changing in a periodic manner (while detailed balance is still satisfied at any time moment) a system may exhibit a nonzero net flux. If the change is adiabatic (slow), then the net flux does not depend on the speed with which the parameters are changed and determined only by the trajectory in parameter space. While the analogy is clear, there are the following differences. In order to have a non-zero additive eigenvalue we postulate that the phase increment may happen whenever the system returns to a previously visited state. The parameters are kept constant. An optimal coordinate is a multi-valued function \textit{per se}, not due to a periodic evolution of parameters. The geometric phase and the net flux are completely determined by the trajectory in the parameter space. The equations for the optimal coordinate are more flexible, they just specify that the optimal coordinate is a multi-valued function, without specifying the exact details; any solution with non-zero $\nu$ can be used.

The solutions presented in the illustrative examples represent a subset of all possible solutions for very simple systems. For example, the equation for the optimal coordinate for a random walk on the line allows other solutions, e.g., solutions with longer periodicity $W_{l+m,i}=W_{l,i}$ for $m>2$, which were not considered. To fully appreciate the properties of the derived equations, it is necessary to fully develop the mathematical formalism similar to the conventional eigenvector decomposition, which would allow one to obtain all the solutions of the equations and to answer the following general questions as the completeness and properties of the basis of additive eigenvectors, the definition of orthogonality or a scalar product. The correspondence between the additive and multiplicative eigenvectors could be useful as a guiding principle. Two other generic questions for the method are obvious. What
are the microscopic models for other relativistic equations of physics and which virtual operators correspond to various stochastic master equations?

It seems that while the solutions using relativistic or symmetric optimal coordinate
are closer to the conventional physical picture, the solutions using left and right eigenvectors are more flexible. Consider for example a random walk in many dimensions. To describe it one can partition the configuration space and compute a transition matrix. One can expect that while at very short time intervals the description by the transition matrix may deviate from the actual dynamics, it will closely approximate it at longer time intervals, when the fine-grained structure of the partitioning can be neglected. The partitioning can be done in many ways, provided that it is sufficiently fine-grained. Thus if one can use sufficiently long time intervals the description of dynamics should be independent of the chosen partition. In particular, if a system performs a random walk along the edges of a cubic lattice, one should be able to accurately describe the dynamics by using any other lattice, i.e., at longer time intervals the space becomes isotropic. The description with the relativistic coordinate, however, is exact only in the limit of $\Delta t\rightarrow 0$, when the original anisotropy of space partitioning is evident.

The fact that we were able to derive model relativistic equations can be explained as follows. The description of dynamics using Markov state models with the master equations is manifestly invariant with respect to the choice of spatial coordinates, since the states are defined only by an index. The new method allows one to reconstruct time, meaning now the temporal coordinate can also be represented just by an index and the description becomes invariant with respect to the choice of spatial-temporal coordinates. 
Such a description can be used to describe dynamics of an arbitrary system using an arbitrary moving frame of reference in an invariant way. By observing a systems trajectory, one can reconstruct time. Having the trajectory as a function of time one can reconstruct the transition probability (or rate) matrix and thus obtain the complete description of the system dynamics. Alternatively, the dynamics can be described by an optimal coordinate, which can be determined directly from the averaged matrix (Eqs. \ref{wlrandt} or \ref{wlrandt2}). To predict a future state of the system as a function of time, one can use an auxiliary system as a clock. Note that the equivalent description which uses two coordinates - the left and right additive eigenvectors, does not exhibit any explicit relativistic effects, in particular, $\nu$ is independent of $\kbar$.

Since the method is capable of reconstructing time intervals from a trajectory without time stamps, it can be applied to "reconstruct time" from inherently timeless objects.

\textbf{In conclusion,} we have suggested a general method for the description of stochastic dynamics. The dynamics is described by using optimal coordinates or \textit{additive} eigenvectors. While, the mathematical formalism is not yet developed to completely characterize the solution space, we believe that we have demonstrated the self-consistency of the method and its potential. 

\begin{acknowledgments} I am grateful to Polina Banushkina and Dmitri Nechaev for numerous discussions. The work was partially supported by an RCUK fellowship.
\end{acknowledgments}

\bibliographystyle{apsrev}

\end{document}